\documentclass[]{aa}
\usepackage{lmodern}
\usepackage{graphicx}
\usepackage{natbib}
\bibpunct{(}{)}{;}{a}{}{,} 
\usepackage{multicol}
\usepackage{multirow}
\usepackage{float}
\usepackage{amsmath}
\usepackage{txfonts}
\usepackage{url}

\makeatletter
\renewcommand*\aa@pageof{, page \thepage{} of \pageref*{LastPage}}
\makeatother

\usepackage[colorlinks=true,allcolors=blue]{hyperref}

\newcommand{\coldsim}{\textsc{ColdSIM}\xspace}

\newcommand{\CI}{\ensuremath{\text{C\,\textsc{i}}}\xspace}
\newcommand{\CII}{\ensuremath{\text{[C\,\textsc{ii}]}}\xspace}
\newcommand{\CIII}{\ensuremath{\text{C\,\textsc{iii}}}\xspace}
\newcommand{\CIV}{\ensuremath{\text{C\,\textsc{iv}}}\xspace}

\newcommand{\HI}{\ensuremath{\text{H\,\textsc{i}}}\xspace}
\newcommand{\HII}{\ensuremath{\text{H\,\textsc{ii}}}\xspace}

\begin{document}
 
\title{\coldsim predictions of \CII emission in primordial galaxies }
\author{Benedetta Casavecchia,\inst{1}
        Umberto Maio,\inst{2,3}
        C\'eline P\'eroux,\inst{4,5}
        Benedetta Ciardi\inst{1}
    }
\institute{
    Max-Planck-Institut f\"ur Astrophysik, Karl-Schwarzschild-Strasse 1, 85748 Garching b. M\"unchen, Germany
    \and
    INAF-Italian National Institute of Astrophysics, Observatory of Trieste, via G. Tiepolo 11, 34143 Trieste, Italy
    \and
    IFPU-Institute for Fundamental Physics of the Universe, Via Beirut 2, 34014 Trieste, Italy
    \and
    European Southern Observatory, Karl-Schwarzschild-Str. 2, 85748 Garching bei München, Germany
    \and
    Aix Marseille Université, CNRS, Laboratoire d’Astrophysique de Marseille (LAM) UMR 7326, 13388 Marseille, France
   }
\offprints{Benedetta Casavecchia, e-mail address: benecasa@mpa-garching.mpg.de}

\titlerunning{\coldsim\ -- \CII emission}
\authorrunning{B. Casavecchia et al.}


\abstract
 {A powerful tool with which to probe the gas content at high redshift is the \CII 158 $\mu$m submillimetre emission line, which, due to its low excitation potential and luminous emission, is considered a possible direct tracer of star forming gas.}
 {In this work, we investigate the origin, evolution, and environmental dependencies of the \CII 158 $\mu$m emission line, as well as its expected correlation with the stellar mass and star formation activity of the high-redshift galaxies observed by JWST.}
 {We use a set of state-of-the-art cold-gas hydrodynamic simulations (\coldsim) with fully coupled time-dependent atomic and molecular non-equilibrium chemistry and self-consistent \CII emission from metal-enriched gas. We accurately track the evolution of \HI, \HII, and H$_2$ in a cosmological context and predict both global and galaxy-based \CII properties.}
 {For the first time, we predict the cosmic mass density evolution of \CII and find that it is in good agreement with new measurements at redshift $z = 6$ from high-resolution optical quasar spectroscopy. 
 We find a correlation between \CII luminosity, $L_{\textup{\CII}}$, and stellar mass, which is consistent with results from ALMA high-redshift large programs. We predict a redshift evolution in the relation between  $L_{\textup{\CII}}$ and the star formation rate (SFR), 
 and provide a fit to relate $L_{\textup{\CII}}$ to SFR, which can be adopted as a more accurate alternative to the currently used linear relation.}
 {Our findings provide physical grounds on which to interpret high-redshift detections in contemporary and future observations, such as the ones performed by ALMA and JWST, and to advance our knowledge of structure formation at early times.}


\maketitle

\section{Introduction}
\label{sect:intro}

The last few years have seen the beginning of a golden era in the search for galaxies in the epoch of reionisation. This has been made possible by a combination of space facilities, such as the \textit{Hubble} Space Telescope (HST) and 8-10 metre diameter ground-based telescopes (e.g. the ESO Very Large Telescope, VLT, and Keck). Dust and metals have been identified in some high-redshift ($z$) objects with the Atacama Large Millimeter/submillimeter Array (ALMA). These datasets are now being supplemented by cutting-edge observations with the \textit{James Webb} Space Telescope (JWST). This observational progress has naturally given rise to a plethora of theoretical models, ranging from analytic and semi-analytic approaches to numerical simulations.

Therefore, it is now possible to detect and interpret the properties of early galaxies during the epoch of reionisation. As star-forming galaxies are  typically bright in the UV due to the emission from young and massive stars, UV luminosity can be used as a primary tracer to quantify their star formation rate (SFR). 
However, part of this radiation is absorbed by interstellar dust and re-emitted at far-infrared (FIR) wavelengths (\citealt{Fixsen98, Dole06, Burgarella13, Whitaker17, Salim20}). For this reason, accurate estimates of the SFR depend on a multi-wavelength analysis ranging from the rest-UV, to the far-infrared (FIR) and redshifted to millimetre (mm) wavelengths, which thus includes both components of star formation; that is, the unobscured emission and the thermal emission from dust \citep{Bethermin20, Khusanova21, Ferrara22, Bowler24}.

As stars are also responsible for gas metal enrichment \citep{Sargent88,Songaila01,Becker15,Codoreanu17,Tumlinson17,Peroux20}, an additional source of information on star formation comes from observations of the metallicity ($Z$) of the gas. The global evolution of the metal mass in the Universe can be captured by the mass density parameter $\Omega$ of single species. Among the possible tracers of $Z$, carbon is one of the most observed at all redshifts, given its high luminosity \citep{Simcoe06, DOdorico13, D'Odorico22, Davies23}. However, it remains challenging to detect the total amount of carbon mass, because this would require that each ionisation state (\CI, \ion{C}{ii}, \CIII, \CIV, etc.) be observed individually given the different ionisation potentials.
Moreover, some of these ions are known to be weak due to their low oscillation strength.
Historically, observational campaigns have therefore focused on estimates of \CIV, which is the most common ion in hot gas and is expected to be the dominant ionisation state at redshifts of $z<5$. Bright quasars have been used as background sources to probe the incidence of metal-rich absorbers along the line of sight, and to quantify the cosmological \CIV mass density parameter, $\Omega_{\textup{\CIV}}$ 
\citep{Songaila01, Pettini03, Boksenberg03, Simcoe06, Simcoe20, DOdorico13, D'Odorico22}. Various authors have used large samples of \CIV absorbers and have reported a significant increase in $\Omega_{\textup{\CIV}}$ towards lower redshifts \citep[see also][]{Becker09, Ryan-Weber09, Simcoe11}. Recently, \cite{Davies23} employed a large number of VLT/X-Shooter high-resolution quasar spectra to measure $\Omega_{\textup{\CIV}}$, finding that it decreases by a factor of 4.8$\pm$2.0 over the $\sim$ 300 Myr interval between $z\sim$ 4.7 and $\sim$ 5.8. These authors interpret this feature as an indication of rapid changes in the gas ionisation state driven by the evolution of the UV background at the end of hydrogen reionisation. 
They also show that C$^+$ is the dominant state at $z>5$ and for the first time report observations of  $\Omega_{{\textup{C}^+}}$. The strength of probing one ionisation state in particular as opposed to global carbon is that, in this way,  the incident ionising background can be directly constrained.

Beyond providing information on the global metal enrichment of the Universe, the \CII emission line in particular has been very valuable for studying high-redshift galaxies. Ultradeep observations with HST and JWST are now detecting the UV stellar light in galaxies up to $z \gtrsim $ 10 \citep{Bouwens19, Bowler20, Leethochawalit22b, Leethochawalit22a, Donnan23, Donnan24, Harikane23}. In this context, redshifted mm observations provide an important means to both confirm the high-redshift candidates and determine their physical properties 
\citep{Fujimoto19, Schaerer20, Ferrara22, Sommovigo21, Sommovigo22, Pozzi24, Romano24}.
To this end, the \CII 158 $\mu$m (1900.5 GHz) fine-structure transition ($\rm ^2P_{3/2} \rightarrow {}^2P_{1/2}$) has proven to be key, as it is the brightest ionic emission line in star-forming galaxies. Due to its low excitation potential, \CII emission can be produced in photodissociation regions, neutral atomic gas, and molecular clouds \citep{Sargsyan12, Pineda13, Rigopoulou14, Velusamy14, Cormier15, Glover16, Nordon16, Diaz-Santos17, Fahrion17}. Moreover, this line is not strongly affected by dust absorption and its relation with the SFR has already been reported and calibrated in the nearby Universe \citep{Stacey91, Leech99, Boselli02, Sargsyan12, DeLooze11, DeLooze14, Herrera-Camus15, Zanella18, Madden20}.
For these reasons, \CII emission is an excellent tool for investigating the properties of high-$z$ structures, including their SFR and kinematics, as well as their neutral atomic and molecular gas content \citep{Cicone15, HM17, Padma19, Dessauges-Zavadsky20, Schaerer20, Ginolfi20, Aravena24, Jones24, Kaasinen24, Posses24, Rowland24}.
Recently, this bright line has gained in popularity as ALMA is detecting an increasing number of objects with \CII emission at $ z \gtrsim 4$.
While ALMA is expected to observe the redshifted \CII emission line from $ z>1$, in practice its higher bands (i.e. 8, 9 and 10) at  frequencies of $>$385 GHz are more challenging to schedule, because they require lower precipitable water vapour conditions, which rarely occur. In addition, observations with these frequencies offer a smaller primary beam or field of view, which limits the number of objects that can be detected in a given exposure time and therefore decreases the survey efficiency at the telescope.
For this reason, most of the available surveys are done at $z>4$, making ALMA an ideal instrument to observe high-$z$ cold gas. Recently, two major ALMA campaigns observed selected UV-bright galaxies at $z \sim 4-8$. The ALPINE survey \citep{LaFevre20} reported over 100 galaxies in the range of $z=4.5-6$, while the REBELS survey \citep{Bouwens22} identified dozens of galaxies at $z=6.5-7.7$. 
Some works found a clear correlation between $L_{\textup{\CII}}$ and SFR \citep{Dessauges-Zavadsky20, Schaerer20, Fujimoto21, Ferrara22, Schouws22}, while others found a \CII luminosity below the  $L_{\textup{\CII}}$--SFR linear relation
\citep{Maiolino15, Pentericci16, Laporte19, Carniani20, Fujimoto24}. 
However, the uncertainties are large at such high redshifts  and the $L_{\textup{\CII}}$ versus SFR relationship \citep{Harikane20, Romano22} is calibrated only for the most massive galaxies, making it inaccurate in the lower SFR regime.
The CONCERTO instrument \citep{CONCERTO20, Gkogkou23} on the Atacama Pathfinder Experiment telescope (APEX) and purpose-built facilities such as the Atacama Large Aperture Submillimeter Telescope (AtLAST) \citep{Klaassen20} will provide further samples of \CII detections in the future.

Theoretical models play a central role in elucidating which mechanisms influence the emission of C$^+$.
Three main approaches have been proposed to derive the \CII luminosity. Semi-analytical models (see e.g. \citealt{Lagache18}, \citealt{Popping19}) use empirical relations to quantify $L_{\textup{\CII}}$ in a cosmological context. The strength of these methods is the fast runtime of the calculations, which leads to the possibility of exploring a large parameter space. These models are also implemented in order to explain single-halo properties, such as the extended ($\sim 10\,\rm kpc$) \CII emission around galaxies at $z = 6$ generated by supernova-driven cooling outflows \citep{Pizzati20, Pizzati23}.
Zoom-in simulations of a limited number of halos are able to capture the small-scale physics at play within a cosmological framework (see e.g. \citealt{Katz19}, \citealt{Bisbas22}, \citealt{Lahen24} and \citealt{Schimek24}).
Finally, simulations run in cosmological boxes offer a complementary avenue, which provides much larger statistics while remaining somewhat limited in terms of mass resolution, in particular for the cold gas (see e.g. \citealt{Leung20}, \citealt{Kannan22}, \citealt{Garcia23}, \citealt{Katz23}, \citealt{Liang24}).
Importantly, we stress that modelling below the hydrodynamic simulation resolution (`subgrid' modelling) is required to address this phase of the gas in both zoom-in and cosmological simulations.
In this respect, simple analytic techniques, for example those based on either gas metallicity or pressure
\citep[][]{Blitz06, Krumholz13}, are often used to split neutral hydrogen into its atomic and molecular components and infer possible \CII properties \citep{Lagos15, Popping19,  Popping22, Szakacs22, Vizgan22a, Vizgan22b}.
Alternatively, different authors have suggested either idealised models of \CII\ line emission \citep{Olsen17}, or post-processing of zoom-in simulations; these latter, although lacking the \CII treatment at runtime, have provided some interesting information on the resolved properties of specific individual haloes \citep{Pallottini19}.
Simulating the atomic and  molecular phases of the cold gas in a full cosmological context, coupled to a consistent implementation for \CII emission, will therefore be one of the most crucial and challenging objectives of studies of galaxy formation and evolution in the decades to come.

The goal of this work is to make a step forward in capturing the complex physical processes at play by running and analysing a set of the state-of-the-art hydrodynamic cosmological simulations, \coldsim \citep{Maio22}, which include fully coupled time-dependent atomic and molecular non-equilibrium chemistry, and lead to self-consistent calculations of \CII emission.
By extending our previous work on primordial molecular gas, we focus here on the \CII line emission in the early Universe and its correlations with the physical properties of the host halo.
Specifically, we study the evolution of the C$^+$ mass density parameter, $\Omega_{{\textup{C}^+}}$, and the dependence of \CII luminosity, $L_{\textup{\CII}}$, on stellar mass, $ M_{\star}$, and SFR in the redshift range $z\sim 6-12$.

The paper is organised as follows:
Section~\ref{sect:methods}
presents the hydrodynamical cosmological simulations used in this study, as well as a detailed description of the methodology to compute the \CII luminosity. 
Section~\ref{sect:Results} details our analysis and associated results. We discuss the impact of our findings 
in section~\ref{sect:discussion},  
and present our conclusions in Section~\ref{sect:conclusions}. 
We adopt a standard cosmology with cold dark matter and cosmological constant $\Lambda$. 
The expansion parameter is 
$H_0$ = 70 km\,s$^{-1}$\,Mpc$^{-1}$ while baryon, matter, and $\Lambda$ density parameters are $\Omega_{\rm b,0}$ = 0.045, $\Omega_{\rm m,0}$ = 0.274, and $\Omega_{\Lambda,0}$ = 0.726, respectively.
We adopt the notation cMpc to indicate lengths in comoving megaparsecs and,  unless otherwise stated, express masses and metallicities in solar units (M$_\odot$ and Z$_\odot$, respectively).

\section{Methods}
\label{sect:methods}

In this section, we introduce the simulations used for our analysis, as well as the methodology adopted to calculate the amount of \CII produced and the associated luminosity.

\subsection{ The \coldsim numerical simulations }
The set of cosmological hydrodynamical simulations used throughout this work is based on the larger \coldsim project \citep{Maio22}.
The latter explores the formation of galaxies at high redshift by employing a physics- and chemical-rich implementation of the evolution of atomic and molecular gas in primordial environments, and represents a remarkable extension of our previous works (\citealp{Maio07, Maio10, Maio11, Maio16}, \citealp{Maio13, Maio15, Ma17a, Ma17b}).
In addition, the ColdSIM simulations have been successfully validated against available observational constraints and recent determinations of the molecular content of $ z > 7 $ galaxies (see \citealt{Feruglio23}).
Their implementation relies on an extended version of the parallel code P-Gadget3 \citep{Springel05}, which, in addition to gravity and smoothed particle hydrodynamics (SPH), includes an {ad hoc} time-dependent non-equilibrium network of first-order differential equations to solve the processes of ionisation, dissociation, and recombination \citep{Abel97,Yoshida03,Maio07} for the following species: e$^{-}$, H, H$^{+}$, H$^{-}$, He, He$^{+}$, He$^{++}$, H$_{2}$, H$^{+}_{2}$ , D, D$^{+}$, HD, and HeH$^{+}$.
In addition to pristine-gas chemistry, ou implementation also includes metal radiative losses from resonant and fine-structure transitions in metal-enriched gas, such as \CII 158 $\mu$m emission.
Here, we use the same chemical and physical implementation as in \cite{Maio22} with slight differences related to assumptions about some  model parameters, box size, and the resolution of the runs (see details below).
In the following, we briefly summarise the characteristics of the \coldsim\ runs analysed in this work and refer the reader to the original paper for more details.

The primordial chemistry network includes H$_2$ formation through  H$^-$ catalysis ($\textup{H} + \textup{e}^{-} \rightarrow \textup{H}^- + \gamma $ and $\textup{H}^- + \textup{H} \rightarrow \textup{H}_2 + \textup{e}^-$), H$_2^+$ catalysis ($\textup{H} + \textup{H}^{+} \rightarrow \textup{H}^+_2 + \gamma $ and $\textup{H}^+_2 + \textup{H} \rightarrow \textup{H}_2 + \textup{H}^+$), and three-body interaction ($3\textup{H} \rightarrow \textup{H}_2 + \textup{H}$).
In gas enriched by stellar feedback, dust-grain catalysis is also accounted for. In detail, H$_2$ dust-grain catalysis is coupled to the non-equilibrium network and followed for different gas temperatures and metallicities by assuming a $Z$-dependent dust-to-metal ratio and a grey-body dust emission with index $\beta = 2$.
This allows us to estimate the redshift evolution expected for dust-grain temperature in agreement with recent ALMA constraints \citep{daCunha21}.

The chemical network is coupled with the evolution of the heavy elements He, C, N, O, Ne, Mg, Si, S, Ca, and Fe, which are traced individually and spread by stellar feedback through asymptotic giant branch winds and Type Ia  and Type II supernova explosions \citep{Maio10,Maio16}. Collisionless star particles are formed stochastically at each integration time according to the star formation model by \cite{Springel03}, which has been slightly modified to follow molecular runaway cooling.
Stellar particles are considered as a single stellar population with a Salpeter initial mass function (IMF). Chemical enrichment from stellar particles is implemented through SPH kernel smoothing. A UV background is turned on at $z \simeq 6$ using the standard \cite{HaardtMadau96} prescription.
Additional physical processes, such as HI and H$_2$ self-shielding, the photoelectric effect, and cosmic-ray heating are also implemented, as they are likely to play an important role in the cooling of molecular gas \citep{Habing68, Draine87, Bakes94}.

\begin{table}[]

\caption{Simulations characteristics}
\begin{tabular}{|c|c|c|c|c|}
\hline
Name         & $L$   & $N_{\rm part}$          & $m_{\rm DM}$       & $m_{\rm gas}$                          \\
& [cMpc/$h$] & & [${\rm M_{\odot}}/h$]&[${\rm M_{\odot}}/h$]\\ 
\hline
CDM Ref & 10 & $2\times 512^3$  & $9.5 \times 10^5$       & $4.7 \times 10^4$        \\
CDM HR       & 10 & $2\times 1000^3$ & $6.3 \times 10^4$ & $1.3 \times 10^4$ \\
CDM LB       & 50 & $2\times 1000^3$ & $7.9\times 10^6$      & $1.6 \times 10^6$  \\
\hline
\end{tabular}
\tablefoot{From left to right, the columns refer to the name of the simulation, the box side $L$, the number of particles $N_{\rm part}$, the mass of DM particles $m_{\rm DM}$, and the initial mass of gas particles $m_{\rm gas}$.}
\label{tab:simul}
\end{table}

As mentioned, we considered three simulations; these explore different box sizes and mass resolutions to investigate how the physical and chemical processes described above affect the properties of galaxies, particularly their atomic and molecular budgets and the resulting \CII emission at various scales.
The reference run, dubbed CDM Ref, has a box size of $10$ cMpc/$h$ with $2 \times 512^3$ particles evenly divided between gas and dark matter. The simulation with higher resolution (CDM HR) has the same box size, but the cosmic field is sampled  with $2 \times 1000^3$ particles. Finally, we employed a larger box (CDM LB) of $50$ cMpc/$h$ with $2 \times 1000^3$ particles to account for rarer, larger structures that would be missing in the other runs.
The characteristics of these simulations are listed in Table \ref{tab:simul}.
In each cosmological volume, cosmic structures have been identified via a friends-of-friends algorithm and a substructure finder. Only objects with at least 100 total particles were considered. The detailed chemical properties of each simulated halo were retrieved in post-processing by matching the halo-constituting particles and the snapshot particles.

\begin{figure*}[h]
\begin{center}
    \includegraphics[width=0.55 \textwidth]{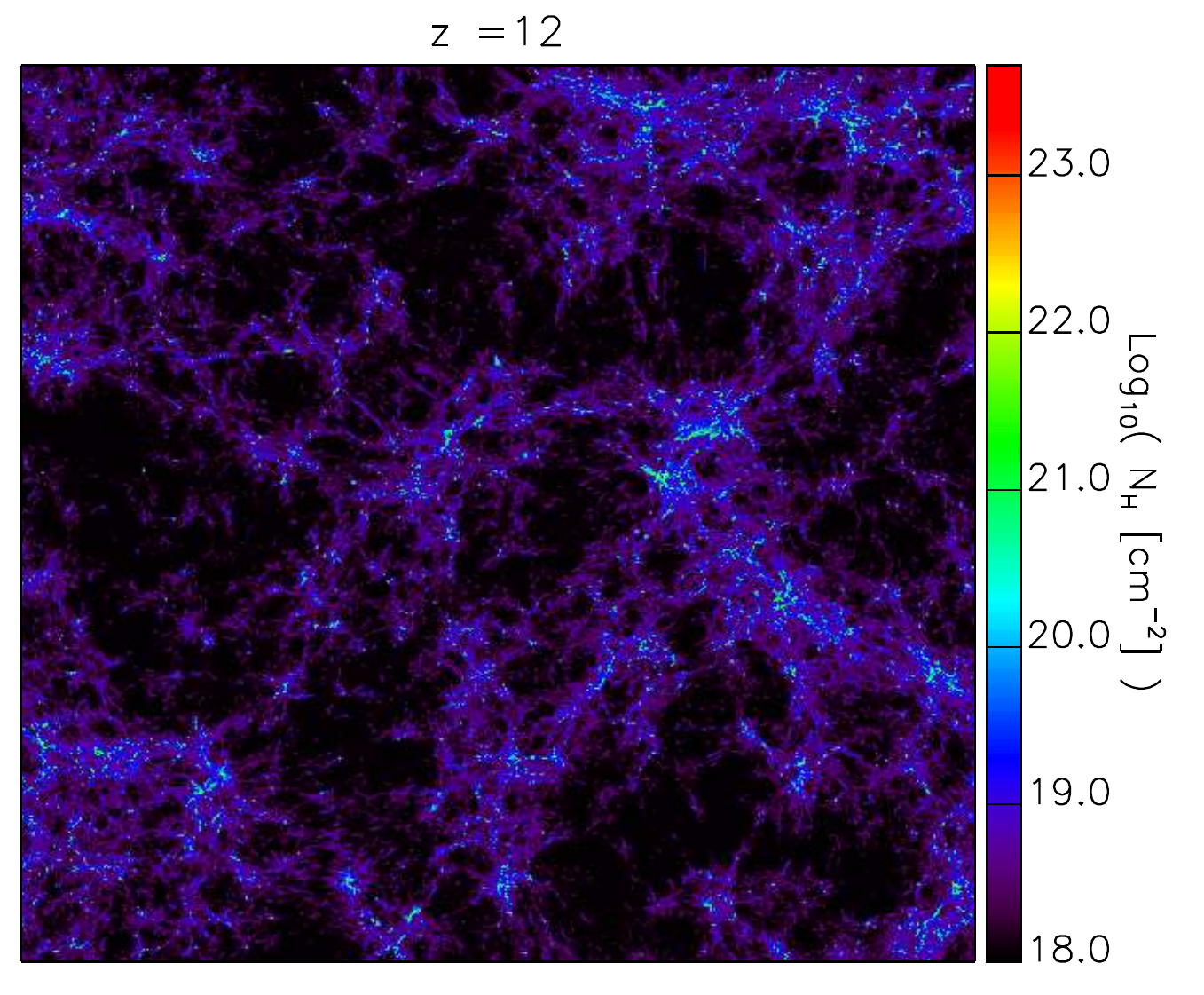}  \hspace{-2.2cm}
    \includegraphics[width=0.55 \textwidth]{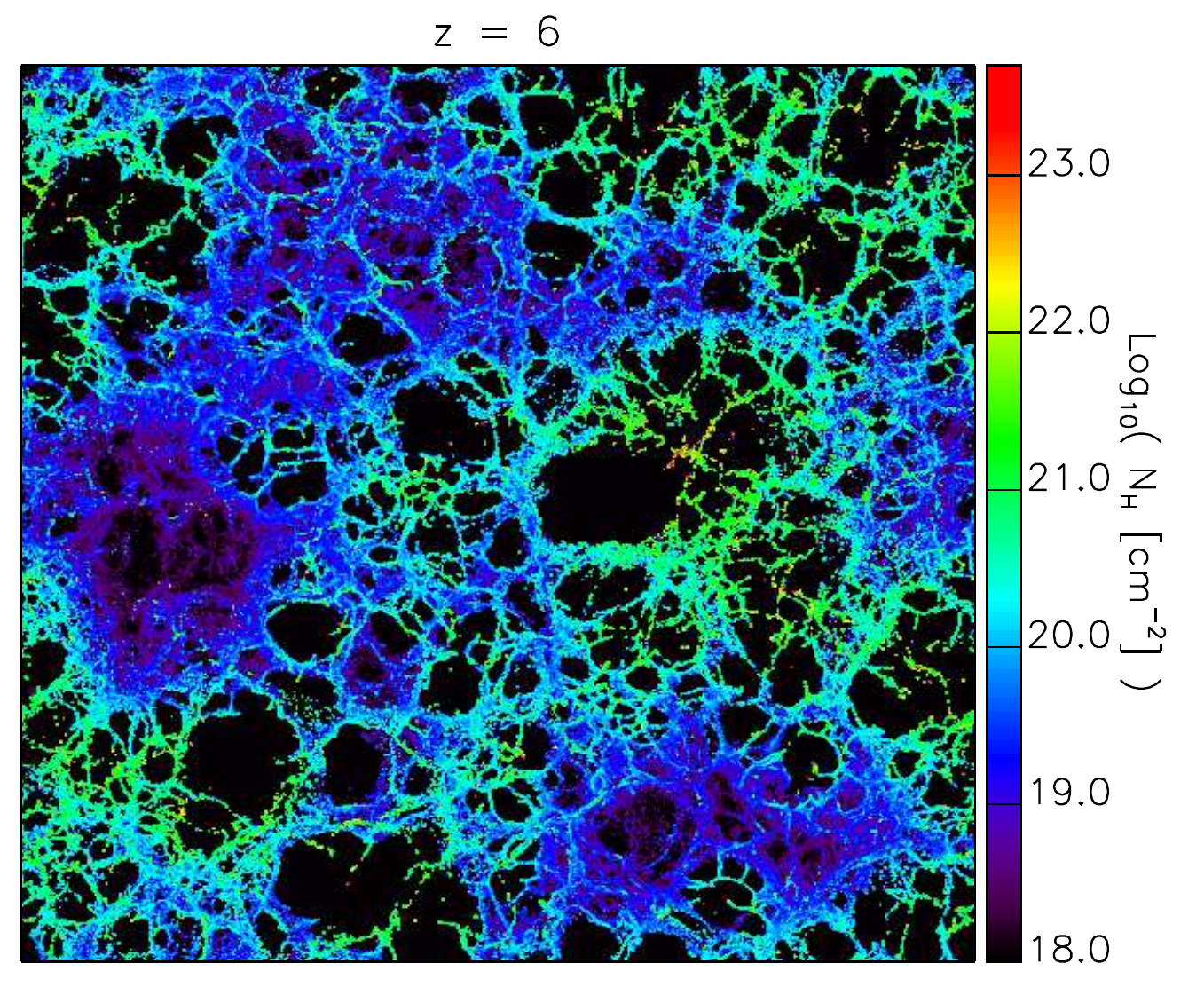}\\
    \includegraphics[width=0.55 \textwidth]{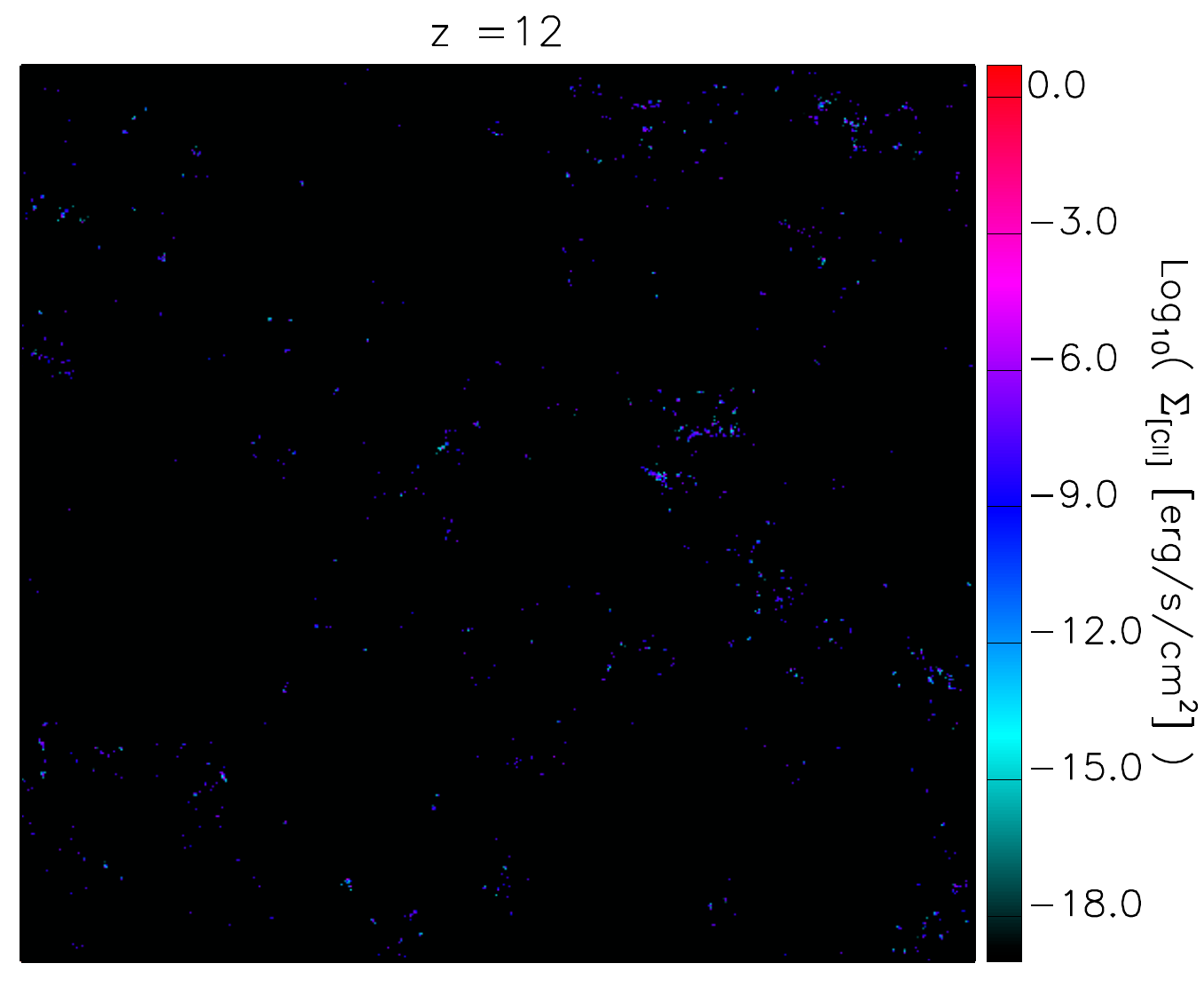} 
    \hspace{-2.2cm}
    \includegraphics[width=0.55 \textwidth]{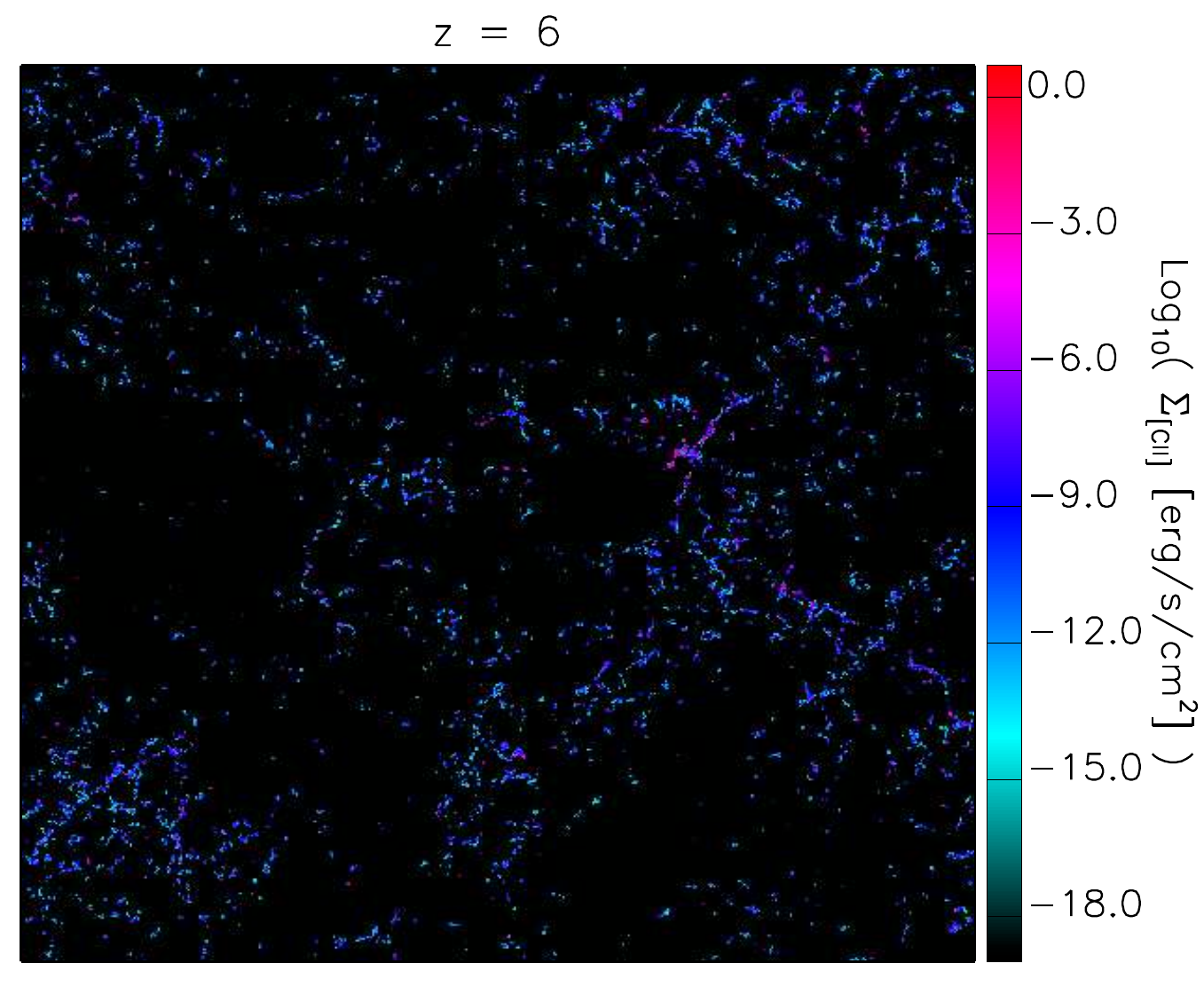}\\
\end{center}
\caption{
   CDM HR column density maps and \CII\ surface-brightness.
   {\it Top row}.
   Column density maps at $ z = 12$ and $ z = 6$ obtained by projecting a slice passing through the centre of the CDM HR simulation box along a width equal to one-tenth of the box side (10 cMpc/$h$) and discretising it on a grid of 512$\times$512 pixels. This results in a pixel linear size of about 19.5~ckpc/{\it h}.
  {\it Bottom row}. \CII surface-brightness maps at $ z = 12$ and $ z = 6$ obtained by projecting the \CII emission from the same slice passing through the centre of the CDM HR simulation box.
  }
  \label{f:maps}
\end{figure*}
As an example, the top row of Fig.~\ref{f:maps} displays gas column density maps at $ z =$ 12 and 6 obtained by projecting a slice of the high-resolution box along its orthogonal dimension.
Cosmological evolution shapes the cosmic medium in dense and void regions (visible in the maps) through the balance between heating and cooling processes. This latter leads to gas collapse and galaxy formation in the densest filamentary environments, as well as metal pollution on stellar timescales.
The injection of carbon atoms ---synthesized by stellar interiors and tracked by our modelling---  into the surrounding medium is then responsible for the \CII emission signal displayed in the bottom row 
(see further discussion below).

\begin{figure}[h]
\begin{center}
  \begin{tabular}{c}
    \includegraphics[width=0.45 \textwidth]{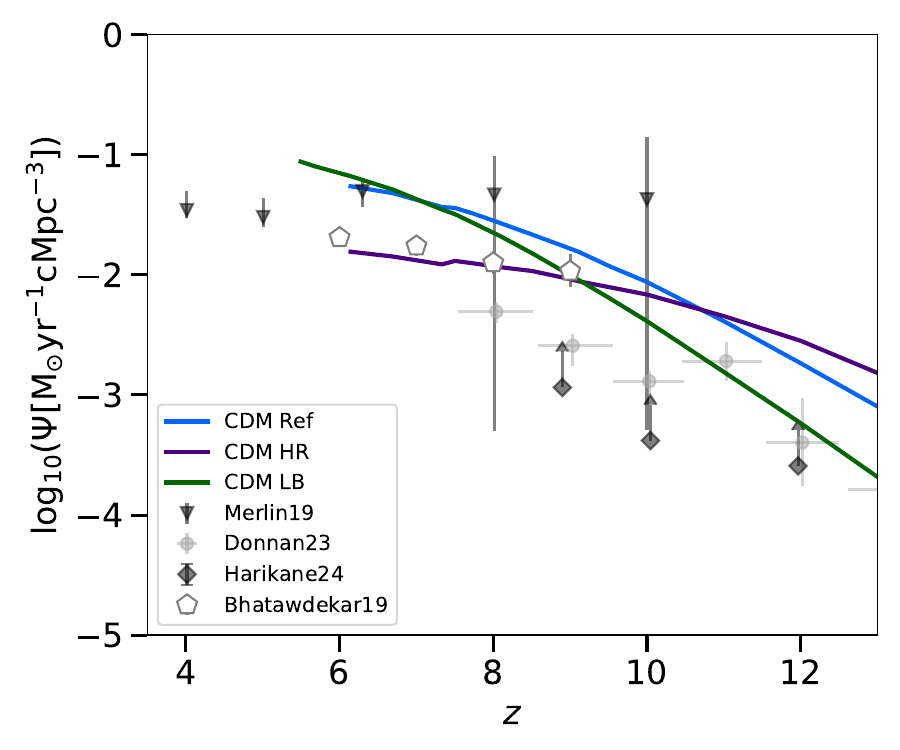}
  \end{tabular}
  \caption{Redshift evolution of the SFR density, $\Psi$, for CDM HR (purple line), CDM Ref (blue), and CDM LB (green). Data points are obtained from NIR (\citealt{Merlin19}, triangles, and \citealt{Bhatawdekar19}, pentagons), dust-corrected UV (\citealt{Donnan23}; circles), and JWST spectroscopic data (\citealt{Harikane24}; diamonds). The latter are lower limits.} 
  \label{f:psi}
\end{center}
\end{figure}

In Figure \ref{f:psi} we show the SFR density, $\Psi$, obtained from the simulations, together with observations based on near-infrared (NIR; \citealt{Merlin19, Bhatawdekar19}), dust-corrected UV \citep{Harikane22, Donnan23} data, as well as recent JWST spectroscopic data \citep{Harikane24}. We note that observational samples vary in the way the SFR is calculated, adding important uncertainties to the comparison with simulated data. Additionally,
NIR estimates are more accurate than UV measurements because they are less prone to dust effects. 
At the highest redshifts, the results from CDM HR (purple line) are above those from the other simulations. The higher resolution of CDM HR leads to an earlier collapse of gas and consequent star formation. Similarly, stellar feedback becomes effective in quenching star formation at earlier times, meaning that the trend is reversed at lower redshifts. 
We find that our predictions reproduce the observed trend at $z \geq$ 6, although they lie above the UV estimates. Nevertheless, we note that the more reliable estimates from NIR data \cite[e.g.][]{Merlin19,Bhatawdekar19} give higher values of $\Psi$, albeit in some cases with large error bars.

\begin{figure*}[h]
\begin{center}
  \begin{tabular}{c}
    \includegraphics[width=1\textwidth]{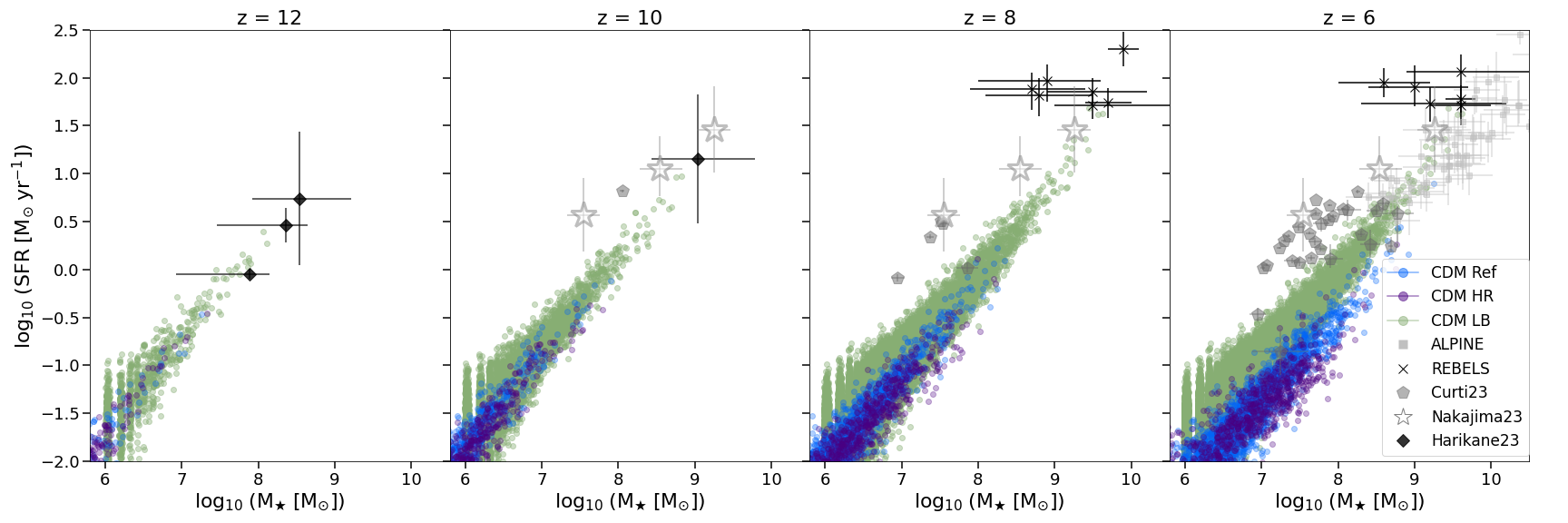}
  \end{tabular}
  \caption{Star formation rate as a function of stellar mass at $z = 12,10,8,$ and 6. The coloured dots refer to predictions from CDM HR (purple), CDM Ref (blue), and CDM LB (green). Grey symbols are data from ALPINE (squares; \citealt{Bethermin20, LaFevre20}), REBELS (crosses; \citealt{Bouwens22,Ferrara22}), and JWST results described in \citet[][stars]{Nakajima23}, \citet[][pentagons]{Curti23}, and \citet[][diamonds]{Harikane23}. Globally, our results agree well with observations.} 
  \label{f:sfrvsmstar}
\end{center}
\end{figure*}

Figure \ref{f:sfrvsmstar} shows the galaxy SFR as a function of stellar mass $  M_\bigstar $ at different redshifts. We find a strong correlation between the stellar mass of galaxies and their SFR for sources on the galaxy main sequence, and so studying its evolution at different redshifts informs us about the efficiency with which galaxies transform gas into stars. We compare our results to a compilation of data from the ALMA large programs ALPINE \citep{Bethermin20, LaFevre20} and REBELS \citep{Bouwens22, Ferrara22}, as well as to JWST results from \cite{Nakajima23} regarding the ERO, GLASS, and CEERS surveys, from \cite{Curti23} for the JADES survey, and to the few detections at $z > 10$ reported by \cite{Harikane23}. Globally, our SFR--$  M_\bigstar $ relation agrees well with the trends of most observations. We note that the predicted SFR of individual haloes is sometimes lower than observed, while in Figure \ref{f:psi} the SFR density indicates that our predictions are slightly higher than observations. 
This could indicate that the number of haloes per unit volume in the simulations is higher than the observed number, which is likely because we simulate haloes with masses of lower than the present detection limit.
The JWST observations reported by \cite{Curti23} and \cite{Nakajima23} reveal a systematically higher SFR for a given stellar mass. Also, the ALMA-based REBELS observations report SFR values that are higher than predicted, albeit with large error bars. We note that the data from \cite{Nakajima23} are averaged over galaxies spanning the range $z =3-10$, and hence might be biased towards high SFRs, and it is possible that other observational studies are similarly biased. Interestingly, the ALMA-based ALPINE results at $z\simeq$ 6 are in good agreement with our predictions from CDM LB.

\begin{figure*}[h]
\begin{center}
  \begin{tabular}{c}
 \includegraphics[width=1\textwidth]{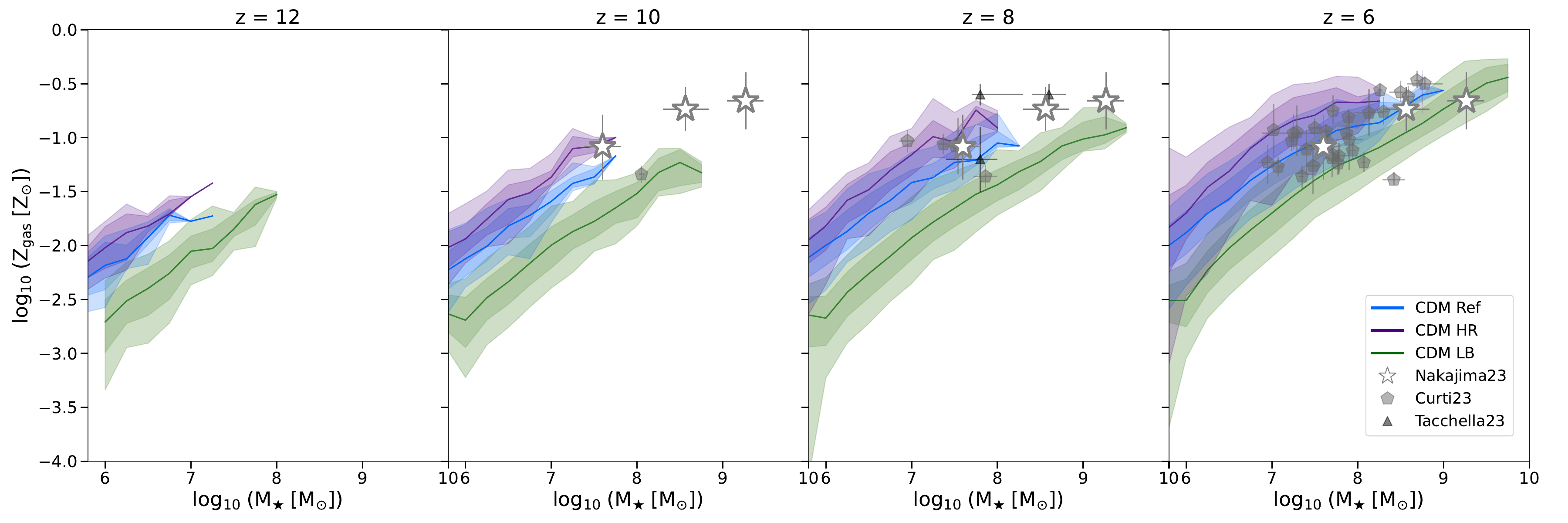}
  \end{tabular}
  \caption{Gas metallicity $Z_{\rm gas}$ as a function of stellar mass $M_\bigstar$ at $z = $ 12, 10, 8, and 6. The solid lines represent the mean values for CDM Ref (blue), CDM HR (purple), and CDM LB (green), while the darker (lighter) shaded regions represent the 1$\sigma$ (2$\sigma$) standard deviation. Grey symbols refer to JWST results described in \citet[][stars]{Nakajima23}, \citet[][pentagons]{Curti23}, and \citet[][triangles]{Tacchella22}. Globally we find that our simulations are in good agreement with JWST observations.} 
  \label{f:zvsmstar}
\end{center}
\end{figure*}

Finally, in Figure \ref{f:zvsmstar} we show the gas metallicity, $Z_{\rm gas}$, as a function of stellar mass at various redshifts. Our results indicate a strong relation among these two quantities. ColdSIM implementation features a non-equilibrium chemical network that includes the individual evolution of heavy elements. Therefore, $Z_{\rm gas}$ is computed for each galaxy as the sum of the masses of the elements heavier than helium within the gas particles divided by the total gas mass. Metal yields are calculated at each time step self-consistently with stellar IMF, lifetimes, and feedback mechanisms.
The metal production from SNe and AGB stars comes naturally from the simulations and is consistent with the stellar evolution model. Non-equilibrium chemistry is tracked while accounting for creation and destruction processes, as well as heating and cooling in the primordial Universe, as described in for example \cite{Maio07, Maio16, Maio22}. In particular, \cite{Maio10} have already shown that carbon and oxygen roughly trace each other at early times.
Therefore, the comparisons made in  Figure~\ref{f:zvsmstar} with oxygen-based JWST metallicities
should be valid (\citealt{Curti23}, \citealt{Nakajima23},  \citealt{Tacchella22}).
The mass--metallicity relation seems to be recovered quite well, with theoretical predictions matching both the trend and the scatter of the observed determinations.
We note that observational estimates of gas metallicities are extremely challenging at these early times and the error bars are very large.
Beyond the relations of the physical galaxy properties reported here, we remind the reader that \cite{Maio23} present additional diagnostics, including UV luminosity functions, halo mass functions, and stellar mass fractions.

\subsection{Galaxy [CII] luminosity}
\label{sect:LCII}

Here we describe how we estimate the galaxy \CII emission at 158 $\mu$m, which is associated with \CII ions excited to the $^{2}P_{3/2}$ level.
The \CII luminosity of a galaxy, $L_{\textup{\CII}}$, is evaluated as the sum of the luminosities of each gas particle, $L_{\textup{\CII},  p}$, contained within it. 
This latter is estimated as 
$L_{\textup{\CII},  p} = \Lambda_{\textup{\CII}} { V_p}$, 
where $ V_p$ is the volume of a sphere with a radius equal to the gas particle smoothing length, and $\Lambda_{\textup{\CII}}$ is the power emitted per unit volume as a consequence of the atomic fine-structure transition.
In the specific case of the \CII 158 $\mu$m emission modelled as a two-level system \citep{Hollenbach89, Goldsmith12, Maio07}, we can write
\begin{equation}
\Lambda_{\textup{\CII}} \equiv 
{n}_{{\textup{C}^+,2}} \, A_{21} \,  \Delta E_{21},
\label{eq:Lambda}
\end{equation}
where $n_{{\textup{C}^+,2}}$ is the number density of the C$^+$ ions excited to the upper state, $A_{21}$ is the Einstein coefficient for spontaneous emission, and $ \Delta E_{21} $ is the energy level separation. The equation above can be rewritten in terms of the total C$^+$ density $ n_{{\textup{C}^+}}$ as \citep[see e.g.][]{Maio07}%
\begin{equation}
\begin{aligned}
\Lambda_{\textup{\CII}} &= \\
&\hspace{-2em} 
\frac{ {n}_{{\textup{C}^+}} \, \left( n_{\rm H_2}\gamma_{12}^{\rm H_2} + n_{\rm H}\gamma_{12}^{\rm H} + n_{\textup{\rm e}}\gamma_{12}^{\rm e} \right) 
} 
{
{ n_{\rm H_2} \left(\gamma_{12}^{\rm H_2} + \gamma_{21}^{\rm H_2} \right) + n_{\textup{\rm H}} \left(\gamma_{12}^{\rm H} + \gamma_{21}^{\rm H} \right) + n_{\textup{\rm e}} \left(\gamma_{12}^{\rm e} + \gamma_{21}^{\rm e} \right) } + A_{21}
}
A_{21} \Delta E_{21},
\end{aligned}
\label{eq:Lambda,1}
\end{equation}
where $\gamma_{12}^{\rm H_2} $, $ \gamma_{12}^{\rm H}$, and $ \gamma_{12}^{\rm e}$ represent the H$_2$-impact, H-impact, and $e^-$-impact collisional excitation rates, $ \gamma_{21}^{\rm H_2}$, $ \gamma_{21}^{\rm H}$, and $ \gamma_{21}^{\rm e}$ denote the rates of de-excitation processes (see the Appendix for specific values), while ${n}_{{\textup{C}^+}}$, $n_{\rm H_2}$, $n_{\textup{\rm H}}$, and $n_{\textup{\rm e}}$ are the number densities of C$^+$, H$_2$, atomic hydrogen, and electrons, respectively.

We note that the resulting \CII emission is a function of temperature, density, and metallicity, and that the emission flux will be especially influenced by the thermal state of the gas below $\sim 10^4$~K, where the coefficient rates of two-body reactions are valid \cite[][]{Maio07, McElroy13}. Above this temperature, carbon is excited at higher excitation levels and does not contribute to \CII emission.

The bottom row of Fig.~\ref{f:maps} shows a pictorial view of the expected \CII surface brightness stemming from a slice of the CDM HR box. For each gas particle, \CII emission is computed according to the actual content, density, and temperature of  carbon according to Eq.~\ref{eq:Lambda,1}. The derived \CII signal in each pixel is given by the integrated emitted power along the line of sight of the pixel divided by the pixel area.
By comparing with the upper panels of Figure~\ref{f:maps}, we observe that \CII emission arises from regions with column densities larger than roughly $ 10^{19}\,\rm cm^{-2}$ and traces metal-enriched parts of the filamentary structures constituting the cosmic medium.
This is in broad agreement with \ion{C}{ii} absorption from damped Lyman-$\alpha$ or subdamped Lyman-$\alpha$ systems \cite[][]{Sebastian24} as well as with preliminary estimates of bright extended \CII haloes at redshift $z\simeq 6$-7 \cite[][]{Pizzati23, Bischetti24}.
Tenuous, diffuse gas that lies far away from galaxy build-up sites is either little enriched with metals (including carbon) or too rarefied to produce significant amounts of \CII surface brightness.
More quantitative findings about features and dependencies of the \CII 158~$\mu$m signal are presented in the following section.\\

\section{Results}
\label{sect:Results}

Here we discuss the results of our simulations in terms of evolution of the cosmological carbon mass density and \CII luminosity functions.

\subsection{Cosmological carbon mass density}

In this section, we analyse the total carbon content and  C$^+$ content in the full cosmological boxes using the corresponding cosmic mass density parameter 
$\Omega_{\rm X} \equiv \rho_{\rm X} / \rho_{\rm crit,0}$, where $\rho_{\rm X}$ is the comoving mass density of element $X$ and $\rho_{\rm crit,0}$ is the present-day critical density.

\begin{figure}[h]
\begin{center}
  \begin{tabular}{c}
    \includegraphics[width=0.45\textwidth]{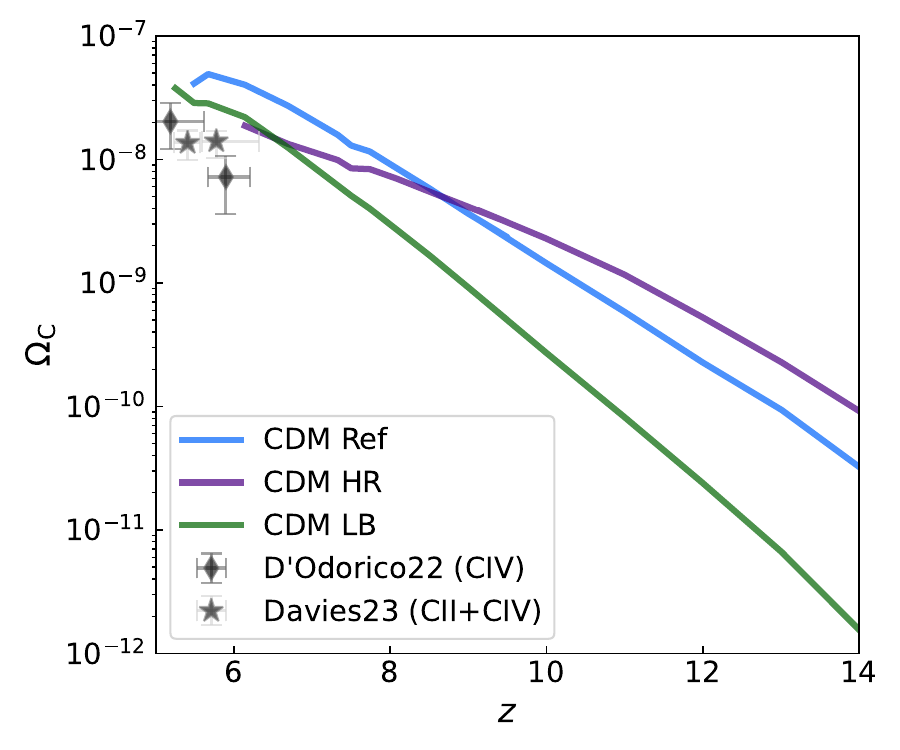}\\
    \includegraphics[width=0.45\textwidth]{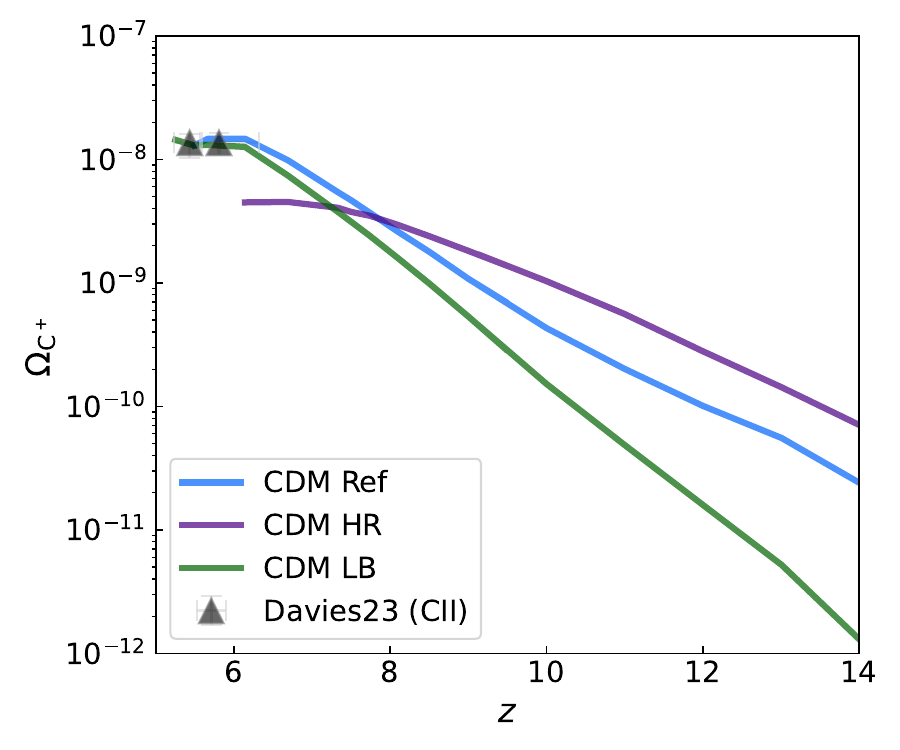}
  \end{tabular}
  \caption{Redshift evolution of the total C mass density, $\Omega_{ \rm C}$ ({\it top panel}), and the C$^+$ mass density, $\Omega_{{\textup{C}^+}}$ ({\it bottom panel}). 
  The lines refer to results from CDM Ref (light blue), CDM HR (purple), and CDM LB (green), while symbols refer to observations of \CIV (diamonds; \citealt{D'Odorico22}), \ion{C}{ii} + \CIV (stars; \citealt{Davies23}), and \ion{C}{ii} (triangles; \citealt{Davies23}). 
  }
  \label{f:omegas}
\end{center}
\end{figure}

While past simulations made predictions as to the cosmological evolution of metal mass densities \cite[e.g.][]{Maio15, Bird16} or \CIV at low redshift \citep{Tescari2011}, here we investigate the evolution of $\Omega_{ \rm C}$ and $\Omega_{{\textup{C}^+}}$ during the epoch of  reionisation, as shown in Figure \ref{f:omegas}.
The upper panel shows that, in all simulations, $\Omega_{ \rm C}$ increases with cosmic time as a natural consequence of the build up of metals. 
At the higher redshifts, the predictions from CDM HR (purple line) lay above those from the other models because its higher resolution leads to an earlier star formation, and hence metal enrichment. For the same reason, in CDM HR, stellar feedback becomes effective in quenching star formation and metal pollution at earlier times, and so the trend is reversed at $z\sim$ 9.5 (6.5) with respect to CDM Ref (CDM LB).

As mentioned in Sect.1, observational estimates of the total C mass density parameter are not available, and thus our results can only be compared to estimates from \ion{C}{ii} and \CIV detections \citep{D'Odorico22, Davies23}, which should be considered as lower limits for $ \Omega_{ \rm C}$. As our simulated values lay above the data, we conclude that our estimates are consistent with observations.

The bottom panel of Figure \ref{f:omegas} shows the evolution of the C$^+$ mass density parameter, which presents features similar to those of $\Omega_{ \rm C}$. Indeed, the flattening of $\Omega_{{\textup{C}^+}}$ here is also related to the earlier effect of stellar feedback in the CDM HR simulation. 
In this case, a direct comparison to observational data \citep{Davies23} is possible and we find excellent agreement with our predictions from CDM LB and CDM Ref, while results from CDM HR are smaller by about a factor of 2.
The fact that $\Omega_{ \rm C}$ is consistent with observations and the differences in $\Omega_{{\textup{C}^+}}$ among simulations are larger than those in $\Omega_{ \rm C}$ at $z < 7$ suggests that such differences are likely not due to chemical evolution but rather to discrepancies in the ionisation state possibly associated to stellar feedback.

\subsection{ {\rm \CII} luminosity function }

In this section, we discuss the \CII luminosity functions obtained from our simulations and compare them to available high-$z$ data.

\subsubsection{Dependence on stellar mass}
\label{sect:lcii-mstar}

\begin{figure*}[h]
\begin{center}
  \begin{tabular}{c}
    \includegraphics[width=1\textwidth]{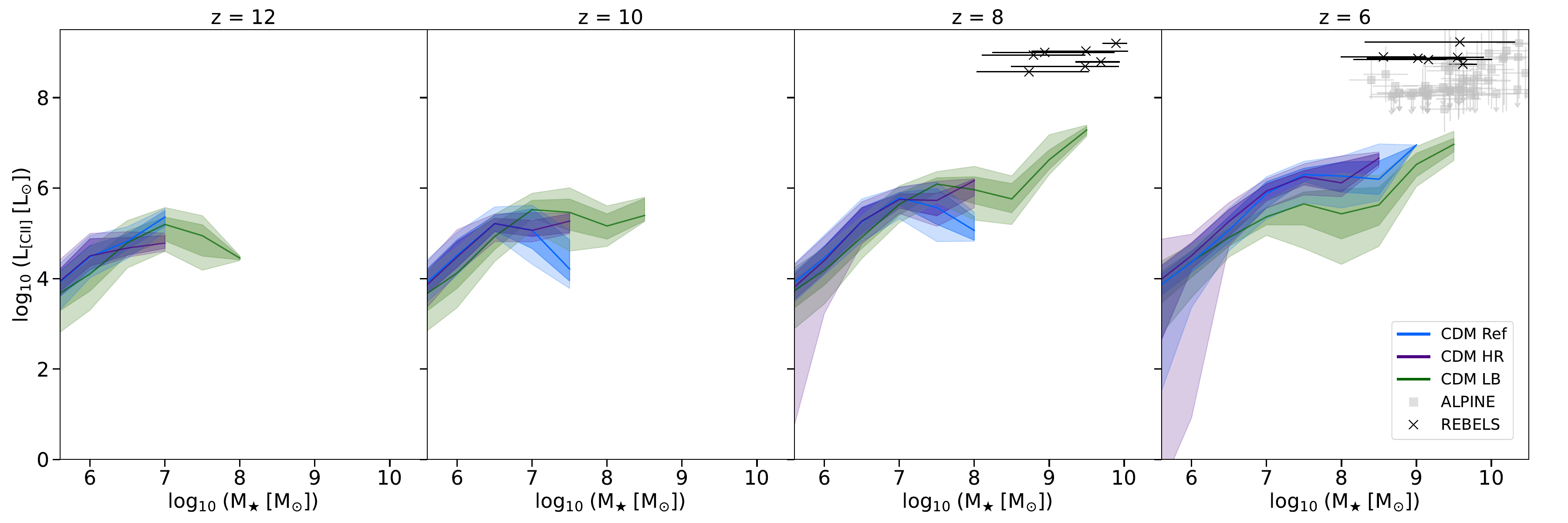}
  \end{tabular}
  \caption{\CII luminosity as a function of halo stellar mass. The panels from left to right display predictions at $z$ = 12, 10, 8, and 6. The solid lines represent the mean values and the darker (lighter) shaded regions represent the 1$\sigma$ (2$\sigma$) standard deviation for CDM Ref (blue), CDM HR (purple), and CDM LB (green). The symbols refer to the ALMA large observational programs ALPINE (squares; \citealt{Bethermin20}, \citealt{LaFevre20}) and REBELS (crosses; \citealt{Bouwens22}, \citealt{Ferrara22}), which target \CII emission at $z=4.5-6$ and $z=6.5-7.7$, respectively. } 
  \label{f:lcii_mstar}
\end{center}
\end{figure*}

In Figure \ref{f:lcii_mstar}, we show $L_{\textup{\CII} }$ as a function of stellar mass, $ M_{\bigstar}$, at different redshifts.  We observe that, typically, more massive galaxies are more luminous in \CII, and that for a given stellar mass, $L_{\textup{\CII}}$ increases with decreasing redshift. This is expected from the build up of stars in galaxies and the ongoing metal enrichment. We also note that in CDM LB we are able to sample bigger objects because of the larger box size. The scatter in these relations is of the order of 1 dex, and increases in the low-SFR regime, especially at stellar masses below $10^{6.5} \;  \rm M_{\odot}$ for CDM HR, which is likely related to the fact that these low-mass haloes have fewer particles. We note a slightly enlarged scatter of 2 dex in the CDM LB, especially for galaxies with stellar masses of $10^{8-9} \;  \rm M_{\odot}$.
We observe that the curves present an irregular trend, with a turnover that becomes visible in the high-mass end, mostly at high redshift. This is discussed further in the following section.

We compare our results with the ALPINE \citep{Bethermin20, LaFevre20} and REBELS \citep{Bouwens22, Ferrara22} ALMA large programs, which target \CII emission at $z=4.5-6$ and $z=6.5-7.7$, respectively. These observations probe rare bright haloes with stellar masses of $\sim 10^9 - 10^{10}$ $ \rm M_{\odot}$ that might even resemble early galaxy groups rather than normal galaxies.
While at $z = 6$ ALPINE detections align well with the trends inferred from the simulated samples, REBELS galaxies have typical luminosities for a given stellar mass that are higher than both the simulated haloes and the ALPINE data.
At such redshifts, there are only a handful of measurements and these are probably biased towards the brightest $L_{\textup{\CII}}$ emitters. 
More precise comparisons will only be possible with future, deeper ALMA observations targeting fainter galaxies.

\subsubsection{Dependence on star formation rate}
\label{results L_CII vs SFR}

\begin{figure*}
\begin{center}
  \begin{tabular}{c}
    \includegraphics[width=1\textwidth]{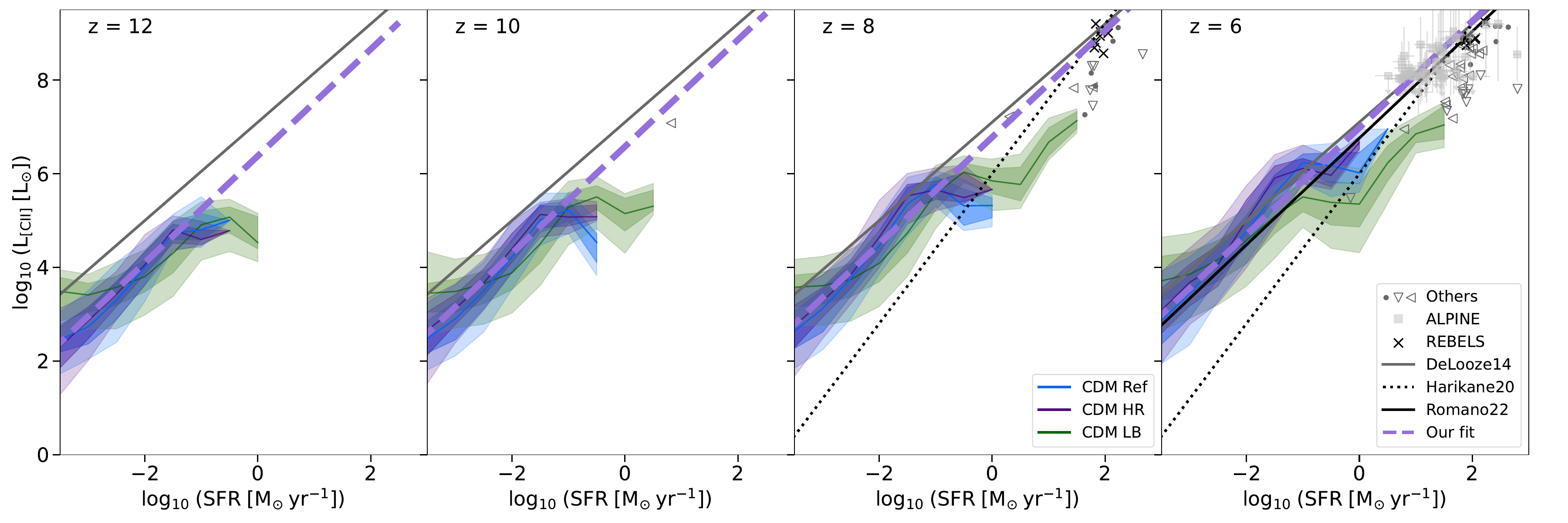}\vspace{-0.1cm}\\
    \includegraphics[width=1\textwidth]{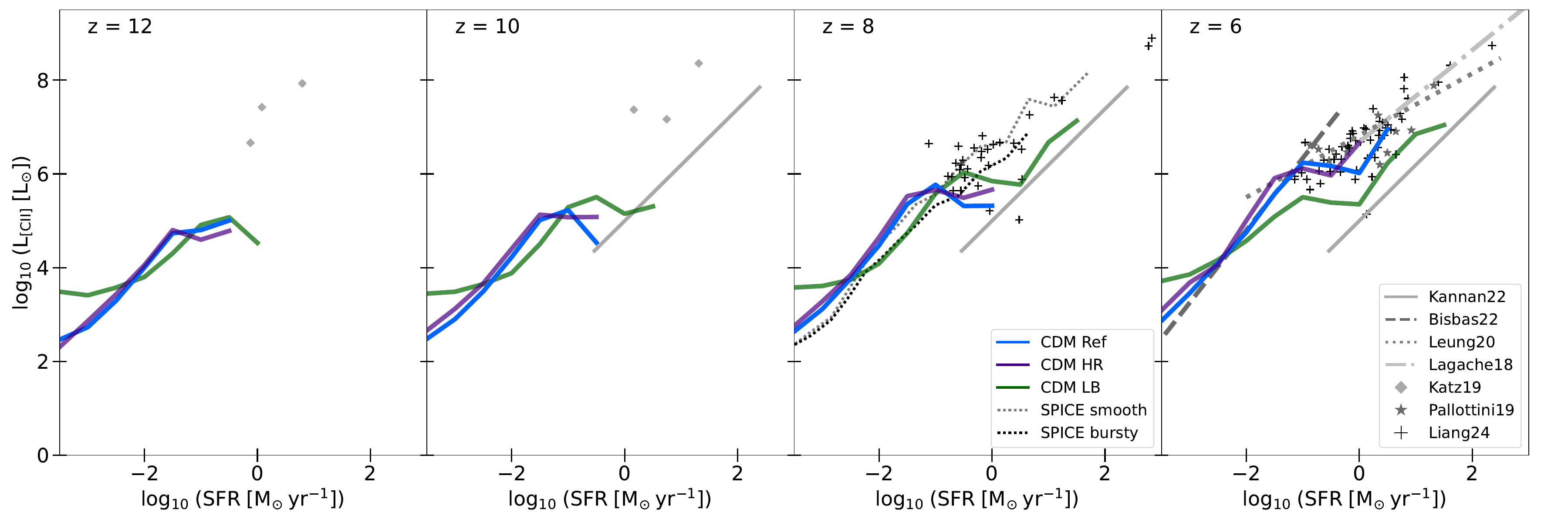}\\
  \end{tabular}
  \caption{\CII luminosity as a function of star formation rate. The top and bottom panels show comparisons with observations and other numerical works, respectively. The panels from left to right display predictions at $z$ = 12, 10, 8, and 6. The solid lines represent the mean values and the darker (lighter) shaded regions represent the 1$\sigma$ (2$\sigma$) standard deviation for CDM Ref (blue), CDM HR (purple), and CDM LB (green). The top row also displays the linear fit to the CDM HR results (purple dashed line), the \cite{DeLooze14} fit made with star forming galaxies in the Local Universe (solid grey), the fit of the ALPINE data ---which accounts for \CII non-detections--- from \citet[][solid black]{Romano22}, and the fit by \cite{Harikane20} to $z = 6-9$ ALMA observed galaxies (dotted black). The symbols refer to the ALMA large observational programs  ALPINE (squares; \citealt{Bethermin20}, \citealt{LaFevre20}) and REBELS (crosses; \citealt{Bouwens22}, \citealt{Ferrara22}), which target \CII emission at $z=4.5-6$ and $z=6.5-7.7$, respectively. The other data are from a compilation from \cite{Liang24}, where downward-pointing triangles refer to upper limits in $L_{\textup{\CII}}$ and left-pointing triangles to upper limits in SFR. In the bottom row, the other grey lines refer to fits of results of previous investigations, while the grey symbols to simulations of individual galaxies. Specifically, we show semi-analytical calculations by \citet[][dashed-dotted line]{Lagache18}, post-processed zoom-in haloes by \citet[][stars]{Pallottini19}, \citet[][diamonds]{Katz19} and \citet[][dashed line]{Bisbas22}, and post-processed cosmological boxes by \citet[][dotted line]{Leung20}, \citet[][solid line]{Kannan22}, \citet[][crosses]{Liang24}, and Bhagwat et al. (in prep.; black dotted and grey dotted lines for the bursty and smooth stellar feedback, respectively). The comparison indicates that the slope and the amplitude of the relation vary from one model to the other, although we note that our results are in good agreement with most observational and theoretical studies reported in this plot.} 
\label{f:lcii_sfr}
\end{center}
\end{figure*}

The top row of Figure \ref{f:lcii_sfr} shows the estimated $L_{\textup{\CII}}$ as a function of SFR at $z$ = 12, 10, 8, and 6. 
Consistently with the previous section, objects with higher SFR are typically more luminous, with a global scatter around the mean value of 1 dex. We note that CDM LB has a slightly larger scatter than the other runs, likely because it contains a greater number of haloes. We stress that there is a redshift evolution in the amplitude of the relation, which increases by about 1 dex from $z=$12 to $z=$6. An important consequence of this result is that, using the observed $L_{\textup{\CII}}$ to estimate the SFR of a galaxy by means of the $L_{\textup{\CII}}$--SFR relation calibrated in the local Universe \cite[e.g.][]{DeLooze14} can lead to inaccurate estimates of high-$z$ SFRs.

At each redshift, we see a turnover in the \CII luminosity for galaxies with higher SFR, as also found in the 
$L_{\textup{\CII}}$--$ M_\bigstar $
behaviour of Section \ref{sect:lcii-mstar}.
We believe that this feature is mainly determined by feedback effects, in particular starbursts, which strongly affect the physical properties of the \CII-emitting gas. A similar feature has indeed been found  by \citealt{Katz23} in simulated `bursty leakers', with SFR averaged over 10 Myr, and also in the radiation hydrodynamic simulations {\tt SPICE} by Bhagwat et al. (in prep.) in their `bursty-sn' model. A less prominent role is additionally played by numerical effects, as suggested by the fact that the feature is more pronounced in CDM LB, because a higher resolution is required to properly model the gas phase that gives rise to \CII emission. 

Although data available at these epochs are sparse and probably biased towards large, bright structures, we highlight that the $L_{\textup{\CII}}$--SFR trend of the simulated galaxies aligns very well with observations.
For a more quantitative comparison, we performed a linear regression fit on the CDM HR sample across the four redshifts. From the fit, we exclude galaxies with SFR$>10^{-1.5}$~$\rm M_{\odot}$~yr$^{-1}$, that is, those affected by the turnover in \CII luminosity. 
Our final relation is

\begin{equation}
    \log_{10}( L_{\textup{\CII}}/[{\rm L_{\odot}}]) = 1.19\, \log_{10} ({\rm SFR/[ M_{\odot} \; yr^{-1}])} - 0.12\, \mathit{z}  + 7.58.
\end{equation}

This is displayed as a purple dashed line in the top panels of Figure \ref{f:lcii_sfr}, and confirms that the extrapolation of our results to higher SFR is in good agreement with available high-$z$ observations. 
We emphasise that, using a single linear relation to infer SFR from $L_{\textup{\CII}}$, as is often done in the literature,
is a relatively crude approximation given the redshift dependence of the relation.
As a reference, in the top row of Figure \ref{f:lcii_sfr} we show the fit to observations of galaxies in the local Universe with SFR $= (10^{-3} - 10^{2})$ $\rm M_{\odot} \: yr^{-1}$ \citep{DeLooze14}.
We also show more recent fits reported in \cite{Harikane20} and \cite{Romano22}, derived from galaxies with SFR $> 10^{0.5}  \: \rm M_{\odot}$.
While at $z=6$ the fit by \cite{DeLooze14} is very similar to ours, as we move to higher $z$ the behaviour of \coldsim galaxies departs from the $L_{\textup{\CII}}$--SFR relation found locally, as a consequence of the mentioned redshift evolution.
The fit by \cite{Harikane20} at $z\simeq 7-8$ is steeper than both the fit by \cite{DeLooze14} and the fit to our simulations, while the fit by \cite{Romano22} obtained from the ALPINE sample ---which also includes \CII non-detections--- is in good agreement with our results.

\section{Discussion}
\label{sect:discussion}

In this paper, we present results from the \coldsim suite of cosmological simulations, focusing in particular on modelling the \CII fine-structure line emission at 158$\,\mu$m.
Previous works have evaluated \CII luminosities in the context of semi-analytical models of structure formation, zoom-in simulations of galaxies, and post-processing of cosmological hydrodynamical boxes.
Among the latter, \coldsim includes a complex time-dependent non-equilibrium chemical network particularly suitable for calculating \CII luminosity, which is expected to arise from cold ($T < 10^4 \,\rm K$) and dense multi-phase gas in its atomic and molecular components.
One of the strengths of the \coldsim simulations is that the evolution of these components, as well as that of C, are computed separately for each gas particle. Thus, the \CII emissivity can be evaluated for each particle by taking into account collisions between carbon and molecular, atomic, and ionised gas.
Furthermore, the \CII radiative power enters the balance between the thermal cooling and heating of each gas parcel at runtime, providing consistent results for gas temperature and chemical abundances.

Here we have focused on the analysis of three \coldsim simulations, which include the same modelling and parameters regulating the various physical processes, and differ only in their box size and resolution. 
Indeed, the larger box size employed in CDM LB allows us to model galaxies with SFRs, stellar masses, and luminosities that are closer to the observed ones, while CDM HR is more appropriate for capturing the small-scale physics regulating the properties of the dense cold gas.
We note that despite some differences observed in the evolution of the mass density estimates, the three simulations give consistent results in terms of $L_{\textup{\CII}}$. This is due to the fact that while the global amounts of metals in boxes  with different sizes and resolutions may vary from one run to another, the \CII luminosities converge with physical  halo properties, such as stellar mass and SFR.

Our findings highlight that, while in the literature $L_{\textup{\CII}}$ is often assumed to correlate linearly with SFR, there is an evident redshift evolution of its amplitude that is independent of resolution and box size. Here we provide a more physically motivated $L_{\textup{\CII}}$--SFR relation that can be employed in place of the typically used linear relation. 

As the predictions discussed in this paper are sensitive to the details of the adopted feedback models \citep{Casey14, Narayanan17}, in the bottom row of Figure \ref{f:lcii_sfr} we compare our results for the $L_{\textup\CII}$--SFR relation with those of a number of previous theoretical investigations. We stress that the various models differ, among others, in terms of resolution, box size, feedback, subgrid physics implementation, and $L_{\textup{\CII}}$ post-processing, meaning that a direct, quantitative comparison is not feasible. \cite{Lagache18}, for example, use a semi-analytical approach that includes a chemodynamical model from metal-free primordial accretion to account for the unresolved  physics. Hydrodynamic zoom-in simulations are instead run by \cite{Katz19}, \cite{Bisbas22}, \citet[][\textsc{SERRA} suite]{Pallottini19}, and \citet[][\textsc{massiveFIRE} simulations]{Liang24}. These simulations model the evolution of galaxies with different properties, such as stellar mass, metallicity, and resolution, spanning a range of galactic viral masses that goes from $10^{10}$ M$_{\odot}$ in \cite{Bisbas22} to $10^{13}$ M$_{\odot}$ in \cite{Liang24}. The former simulation is designed to reproduce a single dwarf-galaxy merger with a gas mass particle resolution of $4$ M$_{\odot}$, while the latter simulates the formation and evolution of giant quiescent galaxies.
Great effort was also made to extract \CII luminosities from cosmological volumes, which cover a range in box size from  $\sim 100$~cMpc/$h$  in {\tt SIMBA} \citep{Leung20} down to 10~cMpc/$h$ in {\tt SPICE} \citep{Bhagwat24}, in which a resolution of $\sim 10^3$~M$_{\odot}$ for stellar particles is reached.
We stress that all these efforts only follow the evolution of the global metallicity, except for \cite{Bisbas22} who model individual elements as in the \coldsim approach. Despite the very different implementations employed in the various investigations, we note a general overlap and agreement in the trend between other theoretical works and \coldsim results, except for \cite{Kannan22}, who predict systematically lower values of $L_{\textup{\CII}}$. 

The simulations presented here, as all others in the literature, rely on a number of assumptions and parameters that have been extensively described and discussed in previous works (see e.g. \citealt{Maio22} and \citealt{Maio23}).
Specifically relevant to the calculation of $ \Omega_{\textup{C}^+}$ and $L_{\textup{\CII}}$ are the uncertainties related to carbon stellar yields, as the two quantities correlate with the mass of carbon injected by stellar processes. In \coldsim, individual elements, including carbon, are tracked separately and not inferred from a global metallicity. This means that the amount of carbon of gas and star particles is estimated at each simulation time step according to the underlying stellar evolution model (i.e. consistently with the adopted IMF, stellar lifetimes, mass-dependent metal yields, and feedback mechanisms, which may increase or decrease the actual chemical abundances). These features make our method robust and precise.
Assumptions as to the initial stellar metallicity and the modelling of the explosive nucleosynthesis may result in differences in the carbon yields by up to a factor of two \citep{Wheeler95, Chieffi04, Kobayashi06}.
The major uncertainty is related to the fact that yield calculations are usually based on one-dimensional stellar models, which is because realistic three-dimensional treatments are much more costly in terms of computing time.
The modelling of angular-momentum loss due to stellar winds and of stellar rotation provide additional sources of uncertainty in the computation of C yields, which can amount up to 1 dex for stars of masses of greater than $30$ M$_{\odot}$ \citep{Limongi18}.

As in our implementation we explicitly include the \CII 158 $\mu$m line transition within gas-cooling calculations, we are able to quantify the \CII luminosity of each gas particle as a function of its hydrodynamic temperature and density in a fully consistent manner.
The regime in which \CII emission is the brightest spans the temperature range between about $10^2$ and $10^4 \,\rm K$. However, uncertainties might come from the adopted collisional rates for the interactions between C$^+$ and other species (electrons, H atoms, H$_2$ molecules) and could affect the estimates of \CII-emitted power.
Collisional rates are computed according to standard atomic physics  and are often available in the literature \cite[e.g.][]{Hollenbach89, Santoro06, Maio07}.
Rates of H-impacts (usually the most abundant at temperatures below $10^4\,\rm K$) are generally well determined.
On the contrary, $e^-$-impact rates, albeit a few thousand times higher, are less robust. As the residual electron abundance in cold neutral gas is negligible, this lack of knowledge should not represent a serious problem for our calculations of \CII\ emission.
H$_2$-impact collisional rates are lower than the H-impact ones by more than a factor of two at temperatures of $ \sim 10^2 \,\rm K $, while the typical reservoir of cold-gas H$_2$ is rarely larger than the neutral H abundance \citep{Saintonge17, Calette18, Hunt20, Zanella23, Hagedorn24}.
This means that our results should be unaffected by atomic-physics uncertainties \cite[see e.g.][and references therein]{Blum92, Wilson02, Goldsmith12}.

\section{Conclusions}
\label{sect:conclusions}

\CII  158~$\mu$m line emission from high-$z$ galaxies was recently measured for the first time by facilities such as ALMA \citep{Bethermin20, LaFevre20}.
It has thus become important to have accurate theoretical models of cold low-temperature gas ($T < 10^4$~K)
in order to understand the physical conditions that give rise to the \CII signal.

In this work, we studied the cold-gas properties of high-$z$ galaxies by analysing \coldsim, a set of state-of-the-art numerical simulations  that include gravity and hydrodynamic calculations, time-dependent atomic and molecular non-equilibrium chemistry, gas cooling and heating, star formation, stellar evolution, and feedback effects in a cosmological context \cite[see][for details]{Maio22}.
We have thus been able to track at each time step the evolution of \HII, \HI, H$_2$, and several metal species in the regime where \CII fine-structure line emission takes place.
Our main results can be summarised as follows:

\begin{itemize}
    \item
     \coldsim results in general reproduce the same trends observed by ALMA and JWST with respect to SFR density and SFR--$M_{\bigstar}$ and $Z_{gas}$--$M_{\bigstar}$ relations. \coldsim SFRs of individual haloes reach regimes that are lower than the ones observed by \cite{Bouwens22}, \cite{Curti23}, and \cite{Nakajima23}. 
    \item 
    For the first time, we compute the redshift evolution of the global mass densities of total carbon and C$^+$ in cosmological simulations, finding good agreement between our predictions and recent data from the XQR-30 survey \citep{Davies23}.
    \item
    Our estimates of the dependence of \CII luminosities on galactic properties, such as SFR and stellar mass, are in good agreement with ALPINE and REBELS observations. It should nevertheless be noted that the range of masses and SFRs analysed and/or observed is not always the same and thus the comparison has to rely on extrapolation of the results.
    \item
    The amplitude and the scatter of the relations differ from one redshift to another, indicating a possible evolution in time of the mentioned correlations.
   \item 
   We provide a fitting formula that relates $L_{\textup{\CII}}$ to SFR and redshift, noting that using a constant relation to link $L_{\textup{\CII}}$ to SFR at $z >$ 6, as often done in observational studies, is not a good approximation, because not only does the relation evolve with redshift, but so do its amplitude and scatter.
\end{itemize}

\noindent
Further studies are still required to fully capture the cold-gas physics and its dependence on modelling assumptions.
Nonetheless, this work provides pivotal physical grounds for the interpretation of high-$z$ \CII detection in contemporary and future observations and demonstrates a new way to theoretically investigate the cold gas in primordial times.

\begin{acknowledgements}
We thank the anonymous referee for the constructive report.
We are thankful to Aniket Bhagwat for providing a preview of the results on \CII luminosity in the {\tt SPICE} galaxies and for fruitful discussions on emission processes in high-$z$ galaxies.
UM acknowledges financial support from the theory grant no. 1.05.23.06.13 ``FIRST -- First Galaxies in the Cosmic Dawn and the Epoch of Reionization with High Resolution Numerical Simulations'' and the travel grant no. 1.05.23.04.01 awarded by the Italian National Institute for Astrophysics.
This research was supported by the International Space Science Institute (ISSI) in Bern (Switzerland), through ISSI International Team project no. 564 (The Cosmic Baryon Cycle from Space).
The numerical calculations done throughout this work have been performed on the machines of the Max Planck Computing and Data Facility of the Max Planck Society, Germany.
We acknowledge the NASA Astrophysics Data System for making available their bibliographic research tools. 
\end{acknowledgements}

\bibliographystyle{aa}
\bibliography{main}

\begin{thebibliography}{150}
\expandafter\ifx\csname natexlab\endcsname\relax\def\natexlab#1{#1}\fi

\bibitem[{{Abel} {et~al.}(1997){Abel}, {Anninos}, {Zhang}, \& {Norman}}]{Abel97}
{Abel}, T., {Anninos}, P., {Zhang}, Y., \& {Norman}, M.~L. 1997, \na, 2, 181

\bibitem[{{Aravena} {et~al.}(2024){Aravena}, {Heintz}, {Dessauges-Zavadsky}, {Oesch}, {Algera}, {Bouwens}, {da Cunha}, {Dayal}, {De Looze}, {Ferrara}, {Fudamoto}, {Gonzalez}, {Graziani}, {Hygate}, {Inami}, {Pallottini}, {Schneider}, {Schouws}, {Sommovigo}, {Topping}, {van der Werf}, \& {Palla}}]{Aravena24}
{Aravena}, M., {Heintz}, K., {Dessauges-Zavadsky}, M., {et~al.} 2024, \aap, 682, A24

\bibitem[{{Bakes} \& {Tielens}(1994)}]{Bakes94}
{Bakes}, E.~L.~O. \& {Tielens}, A.~G.~G.~M. 1994, \apj, 427, 822

\bibitem[{{Becker} {et~al.}(2015){Becker}, {Bolton}, \& {Lidz}}]{Becker15}
{Becker}, G.~D., {Bolton}, J.~S., \& {Lidz}, A. 2015, \pasa, 32, e045

\bibitem[{{Becker} {et~al.}(2009){Becker}, {Rauch}, \& {Sargent}}]{Becker09}
{Becker}, G.~D., {Rauch}, M., \& {Sargent}, W. L.~W. 2009, \apj, 698, 1010

\bibitem[{{B{\'e}thermin} {et~al.}(2020){B{\'e}thermin}, {Fudamoto}, {Ginolfi}, {Loiacono}, {Khusanova}, {Capak}, {Cassata}, {Faisst}, {Le F{\`e}vre}, {Schaerer}, {Silverman}, {Yan}, {Amorin}, {Bardelli}, {Boquien}, {Cimatti}, {Davidzon}, {Dessauges-Zavadsky}, {Fujimoto}, {Gruppioni}, {Hathi}, {Ibar}, {Jones}, {Koekemoer}, {Lagache}, {Lemaux}, {Moreau}, {Oesch}, {Pozzi}, {Riechers}, {Talia}, {Toft}, {Vallini}, {Vergani}, {Zamorani}, \& {Zucca}}]{Bethermin20}
{B{\'e}thermin}, M., {Fudamoto}, Y., {Ginolfi}, M., {et~al.} 2020, \aap, 643, A2

\bibitem[{{Bhagwat} {et~al.}(2024){Bhagwat}, {Costa}, {Ciardi}, {Pakmor}, \& {Garaldi}}]{Bhagwat24}
{Bhagwat}, A., {Costa}, T., {Ciardi}, B., {Pakmor}, R., \& {Garaldi}, E. 2024, \mnras, 531, 3406

\bibitem[{{Bhatawdekar} {et~al.}(2019){Bhatawdekar}, {Conselice}, {Margalef-Bentabol}, \& {Duncan}}]{Bhatawdekar19}
{Bhatawdekar}, R., {Conselice}, C.~J., {Margalef-Bentabol}, B., \& {Duncan}, K. 2019, \mnras, 486, 3805

\bibitem[{{Biffi} \& {Maio}(2013)}]{Maio13}
{Biffi}, V. \& {Maio}, U. 2013, \mnras, 436, 1621

\bibitem[{{Bird} {et~al.}(2016){Bird}, {Rubin}, {Suresh}, \& {Hernquist}}]{Bird16}
{Bird}, S., {Rubin}, K. H.~R., {Suresh}, J., \& {Hernquist}, L. 2016, \mnras, 462, 307

\bibitem[{{Bisbas} {et~al.}(2022){Bisbas}, {Walch}, {Naab}, {Lah{\'e}n}, {Herrera-Camus}, {Steinwandel}, {Fotopoulou}, {Hu}, \& {Johansson}}]{Bisbas22}
{Bisbas}, T.~G., {Walch}, S., {Naab}, T., {et~al.} 2022, \apj, 934, 115

\bibitem[{{Bischetti} {et~al.}(2024){Bischetti}, {Choi}, {Fiore}, {Feruglio}, {Carniani}, {D'Odorico}, {Ba{\~n}ados}, {Chen}, {Decarli}, {Gallerani}, {Hlavacek-Larrondo}, {Lai}, {Leighly}, {Mazzucchelli}, {Perreault-Levasseur}, {Tripodi}, {Walter}, {Wang}, {Yang}, {Vittoria Zanchettin}, \& {Zhu}}]{Bischetti24}
{Bischetti}, M., {Choi}, H., {Fiore}, F., {et~al.} 2024, arXiv e-prints, arXiv:2404.12443

\bibitem[{{Blitz} \& {Rosolowsky}(2006)}]{Blitz06}
{Blitz}, L. \& {Rosolowsky}, E. 2006, \apj, 650, 933

\bibitem[{{Blum} \& {Pradhan}(1992)}]{Blum92}
{Blum}, R.~D. \& {Pradhan}, A.~K. 1992, \apjs, 80, 425

\bibitem[{{Boksenberg} {et~al.}(2003){Boksenberg}, {Sargent}, \& {Rauch}}]{Boksenberg03}
{Boksenberg}, A., {Sargent}, W. L.~W., \& {Rauch}, M. 2003, arXiv e-prints [\eprint[arXiv]{astro-ph/0307557}]

\bibitem[{{Boselli} {et~al.}(2002){Boselli}, {Gavazzi}, {Lequeux}, \& {Pierini}}]{Boselli02}
{Boselli}, A., {Gavazzi}, G., {Lequeux}, J., \& {Pierini}, D. 2002, \aap, 385, 454

\bibitem[{{Bouwens} {et~al.}(2022){Bouwens}, {Smit}, {Schouws}, {Stefanon}, {Bowler}, {Endsley}, {Gonzalez}, {Inami}, {Stark}, {Oesch}, {Hodge}, {Aravena}, {da Cunha}, {Dayal}, {de Looze}, {Ferrara}, {Fudamoto}, {Graziani}, {Li}, {Nanayakkara}, {Pallottini}, {Schneider}, {Sommovigo}, {Topping}, {van der Werf}, {Algera}, {Barrufet}, {Hygate}, {Labb{\'e}}, {Riechers}, \& {Witstok}}]{Bouwens22}
{Bouwens}, R.~J., {Smit}, R., {Schouws}, S., {et~al.} 2022, \apj, 931, 160

\bibitem[{{Bouwens} {et~al.}(2019){Bouwens}, {Stefanon}, {Oesch}, {Illingworth}, {Nanayakkara}, {Roberts-Borsani}, {Labb{\'e}}, \& {Smit}}]{Bouwens19}
{Bouwens}, R.~J., {Stefanon}, M., {Oesch}, P.~A., {et~al.} 2019, \apj, 880, 25

\bibitem[{{Bowler} {et~al.}(2024){Bowler}, {Inami}, {Sommovigo}, {Smit}, {Algera}, {Aravena}, {Barrufet}, {Bouwens}, {da Cunha}, {Cullen}, {Dayal}, {De Looze}, {Dunlop}, {Fudamoto}, {Mauerhofer}, {McLure}, {Stefanon}, {Schneider}, {Ferrara}, {Graziani}, {Hodge}, {Nanayakkara}, {Palla}, {Schouws}, {Stark}, \& {van der Werf}}]{Bowler24}
{Bowler}, R.~A.~A., {Inami}, H., {Sommovigo}, L., {et~al.} 2024, \mnras, 527, 5808

\bibitem[{{Bowler} {et~al.}(2020){Bowler}, {Jarvis}, {Dunlop}, {McLure}, {McLeod}, {Adams}, {Milvang-Jensen}, \& {McCracken}}]{Bowler20}
{Bowler}, R.~A.~A., {Jarvis}, M.~J., {Dunlop}, J.~S., {et~al.} 2020, \mnras, 493, 2059

\bibitem[{{Burgarella} {et~al.}(2013){Burgarella}, {Buat}, {Gruppioni}, {Cucciati}, {Heinis}, {Berta}, {B{\'e}thermin}, {Bock}, {Cooray}, {Dunlop}, {Farrah}, {Franceschini}, {Le Floc'h}, {Lutz}, {Magnelli}, {Nordon}, {Oliver}, {Page}, {Popesso}, {Pozzi}, {Riguccini}, {Vaccari}, \& {Viero}}]{Burgarella13}
{Burgarella}, D., {Buat}, V., {Gruppioni}, C., {et~al.} 2013, \aap, 554, A70

\bibitem[{{Calette} {et~al.}(2018){Calette}, {Avila-Reese}, {Rodr{\'\i}guez-Puebla}, {Hern{\'a}ndez-Toledo}, \& {Papastergis}}]{Calette18}
{Calette}, A.~R., {Avila-Reese}, V., {Rodr{\'\i}guez-Puebla}, A., {Hern{\'a}ndez-Toledo}, H., \& {Papastergis}, E. 2018, \rmxaa, 54, 443

\bibitem[{{Carniani} {et~al.}(2020){Carniani}, {Ferrara}, {Maiolino}, {Castellano}, {Gallerani}, {Fontana}, {Kohandel}, {Lupi}, {Pallottini}, {Pentericci}, {Vallini}, \& {Vanzella}}]{Carniani20}
{Carniani}, S., {Ferrara}, A., {Maiolino}, R., {et~al.} 2020, \mnras, 499, 5136

\bibitem[{{Casey} {et~al.}(2014){Casey}, {Narayanan}, \& {Cooray}}]{Casey14}
{Casey}, C.~M., {Narayanan}, D., \& {Cooray}, A. 2014, \physrep, 541, 45

\bibitem[{{Chieffi} \& {Limongi}(2004)}]{Chieffi04}
{Chieffi}, A. \& {Limongi}, M. 2004, \apj, 608, 405

\bibitem[{{Cicone} {et~al.}(2015){Cicone}, {Maiolino}, {Gallerani}, {Neri}, {Ferrara}, {Sturm}, {Fiore}, {Piconcelli}, \& {Feruglio}}]{Cicone15}
{Cicone}, C., {Maiolino}, R., {Gallerani}, S., {et~al.} 2015, \aap, 574, A14

\bibitem[{{Codoreanu} {et~al.}(2017){Codoreanu}, {Ryan-Weber}, {Crighton}, {Becker}, {Pettini}, {Madau}, \& {Venemans}}]{Codoreanu17}
{Codoreanu}, A., {Ryan-Weber}, E.~V., {Crighton}, N. H.~M., {et~al.} 2017, \mnras, 472, 1023

\bibitem[{{CONCERTO Collaboration} {et~al.}(2020){CONCERTO Collaboration}, {Ade}, {Aravena}, {Barria}, {Beelen}, {Benoit}, {B{\'e}thermin}, {Bounmy}, {Bourrion}, {Bres}, {De Breuck}, {Calvo}, {Cao}, {Catalano}, {D{\'e}sert}, {Dur{\'a}n}, {Fasano}, {Fenouillet}, {Garcia}, {Garde}, {Goupy}, {Groppi}, {Hoarau}, {Lagache}, {Lambert}, {Leggeri}, {Levy-Bertrand}, {Mac{\'\i}as-P{\'e}rez}, {Mani}, {Marpaud}, {Mauskopf}, {Monfardini}, {Pisano}, {Ponthieu}, {Prieur}, {Roni}, {Roudier}, {Tourres}, \& {Tucker}}]{CONCERTO20}
{CONCERTO Collaboration}, {Ade}, P., {Aravena}, M., {et~al.} 2020, \aap, 642, A60

\bibitem[{{Cormier} {et~al.}(2015){Cormier}, {Madden}, {Lebouteiller}, {Abel}, {Hony}, {Galliano}, {R{\'e}my-Ruyer}, {Bigiel}, {Baes}, {Boselli}, {Chevance}, {Cooray}, {De Looze}, {Doublier}, {Galametz}, {Hughes}, {Karczewski}, {Lee}, {Lu}, \& {Spinoglio}}]{Cormier15}
{Cormier}, D., {Madden}, S.~C., {Lebouteiller}, V., {et~al.} 2015, \aap, 578, A53

\bibitem[{{Curti} {et~al.}(2023){Curti}, {D'Eugenio}, {Carniani}, {Maiolino}, {Sandles}, {Witstok}, {Baker}, {Bennett}, {Piotrowska}, {Tacchella}, {Charlot}, {Nakajima}, {Maheson}, {Mannucci}, {Amiri}, {Arribas}, {Belfiore}, {Bonaventura}, {Bunker}, {Chevallard}, {Cresci}, {Curtis-Lake}, {Hayden-Pawson}, {Jones}, {Kumari}, {Laseter}, {Looser}, {Marconi}, {Maseda}, {Scholtz}, {Smit}, {{\"U}bler}, \& {Wallace}}]{Curti23}
{Curti}, M., {D'Eugenio}, F., {Carniani}, S., {et~al.} 2023, \mnras, 518, 425

\bibitem[{{da Cunha} {et~al.}(2021){da Cunha}, {Hodge}, {Casey}, {Algera}, {Kaasinen}, {Smail}, {Walter}, {Brandt}, {Dannerbauer}, {Decarli}, {Groves}, {Knudsen}, {Swinbank}, {Weiss}, {van der Werf}, \& {Zavala}}]{daCunha21}
{da Cunha}, E., {Hodge}, J.~A., {Casey}, C.~M., {et~al.} 2021, \apj, 919, 30

\bibitem[{{Davies} {et~al.}(2023){Davies}, {Ryan-Weber}, {D'Odorico}, {Bosman}, {Meyer}, {Becker}, {Cupani}, {Keating}, {Bischetti}, {Davies}, {Eilers}, {Farina}, {Haehnelt}, {Pallottini}, \& {Zhu}}]{Davies23}
{Davies}, R.~L., {Ryan-Weber}, E., {D'Odorico}, V., {et~al.} 2023, \mnras, 521, 314

\bibitem[{{De Looze} {et~al.}(2011){De Looze}, {Baes}, {Bendo}, {Cortese}, \& {Fritz}}]{DeLooze11}
{De Looze}, I., {Baes}, M., {Bendo}, G.~J., {Cortese}, L., \& {Fritz}, J. 2011, \mnras, 416, 2712

\bibitem[{{De Looze} {et~al.}(2014){De Looze}, {Cormier}, {Lebouteiller}, {Madden}, {Baes}, {Bendo}, {Boquien}, {Boselli}, {Clements}, {Cortese}, {Cooray}, {Galametz}, {Galliano}, {Graci{\'a}-Carpio}, {Isaak}, {Karczewski}, {Parkin}, {Pellegrini}, {R{\'e}my-Ruyer}, {Spinoglio}, {Smith}, \& {Sturm}}]{DeLooze14}
{De Looze}, I., {Cormier}, D., {Lebouteiller}, V., {et~al.} 2014, \aap, 568, A62

\bibitem[{{Dessauges-Zavadsky} {et~al.}(2020){Dessauges-Zavadsky}, {Ginolfi}, {Pozzi}, {B{\'e}thermin}, {Le F{\`e}vre}, {Fujimoto}, {Silverman}, {Jones}, {Vallini}, {Schaerer}, {Faisst}, {Khusanova}, {Fudamoto}, {Cassata}, {Loiacono}, {Capak}, {Yan}, {Amorin}, {Bardelli}, {Boquien}, {Cimatti}, {Gruppioni}, {Hathi}, {Ibar}, {Koekemoer}, {Lemaux}, {Narayanan}, {Oesch}, {Rodighiero}, {Romano}, {Talia}, {Toft}, {Vergani}, {Zamorani}, \& {Zucca}}]{Dessauges-Zavadsky20}
{Dessauges-Zavadsky}, M., {Ginolfi}, M., {Pozzi}, F., {et~al.} 2020, \aap, 643, A5

\bibitem[{{D{\'\i}az-Santos} {et~al.}(2017){D{\'\i}az-Santos}, {Armus}, {Charmandaris}, {Lu}, {Stierwalt}, {Stacey}, {Malhotra}, {van der Werf}, {Howell}, {Privon}, {Mazzarella}, {Goldsmith}, {Murphy}, {Barcos-Mu{\~n}oz}, {Linden}, {Inami}, {Larson}, {Evans}, {Appleton}, {Iwasawa}, {Lord}, {Sanders}, \& {Surace}}]{Diaz-Santos17}
{D{\'\i}az-Santos}, T., {Armus}, L., {Charmandaris}, V., {et~al.} 2017, \apj, 846, 32

\bibitem[{{D'Odorico} {et~al.}(2013){D'Odorico}, {Cupani}, {Cristiani}, {Maiolino}, {Molaro}, {Nonino}, {Centuri{\'o}n}, {Cimatti}, {di Serego Alighieri}, {Fiore}, {Fontana}, {Gallerani}, {Giallongo}, {Mannucci}, {Marconi}, {Pentericci}, {Viel}, \& {Vladilo}}]{DOdorico13}
{D'Odorico}, V., {Cupani}, G., {Cristiani}, S., {et~al.} 2013, \mnras, 435, 1198

\bibitem[{{D'Odorico} {et~al.}(2022){D'Odorico}, {Finlator}, {Cristiani}, {Cupani}, {Perrotta}, {Calura}, {C{\`e}nturion}, {Becker}, {Berg}, {Lopez}, {Ellison}, \& {Pomante}}]{D'Odorico22}
{D'Odorico}, V., {Finlator}, K., {Cristiani}, S., {et~al.} 2022, \mnras, 512, 2389

\bibitem[{{Dole} {et~al.}(2006){Dole}, {Lagache}, {Puget}, {Caputi}, {Fern{\'a}ndez-Conde}, {Le Floc'h}, {Papovich}, {P{\'e}rez-Gonz{\'a}lez}, {Rieke}, \& {Blaylock}}]{Dole06}
{Dole}, H., {Lagache}, G., {Puget}, J.~L., {et~al.} 2006, \aap, 451, 417

\bibitem[{{Donnan} {et~al.}(2023){Donnan}, {McLeod}, {Dunlop}, {McLure}, {Carnall}, {Begley}, {Cullen}, {Hamadouche}, {Bowler}, {Magee}, {McCracken}, {Milvang-Jensen}, {Moneti}, \& {Targett}}]{Donnan23}
{Donnan}, C.~T., {McLeod}, D.~J., {Dunlop}, J.~S., {et~al.} 2023, \mnras, 518, 6011

\bibitem[{{Donnan} {et~al.}(2024){Donnan}, {McLure}, {Dunlop}, {McLeod}, {Magee}, {Arellano-C{\'o}rdova}, {Barrufet}, {Begley}, {Bowler}, {Carnall}, {Cullen}, {Ellis}, {Fontana}, {Illingworth}, {Grogin}, {Hamadouche}, {Koekemoer}, {Liu}, {Mason}, {Santini}, \& {Stanton}}]{Donnan24}
{Donnan}, C.~T., {McLure}, R.~J., {Dunlop}, J.~S., {et~al.} 2024, arXiv e-prints, arXiv:2403.03171

\bibitem[{{Draine} \& {Sutin}(1987)}]{Draine87}
{Draine}, B.~T. \& {Sutin}, B. 1987, \apj, 320, 803

\bibitem[{{Fahrion} {et~al.}(2017){Fahrion}, {Cormier}, {Bigiel}, {Hony}, {Abel}, {Cigan}, {Csengeri}, {Graf}, {Lebouteiller}, {Madden}, {Wu}, \& {Young}}]{Fahrion17}
{Fahrion}, K., {Cormier}, D., {Bigiel}, F., {et~al.} 2017, \aap, 599, A9

\bibitem[{{Ferrara} {et~al.}(2022){Ferrara}, {Sommovigo}, {Dayal}, {Pallottini}, {Bouwens}, {Gonzalez}, {Inami}, {Smit}, {Bowler}, {Endsley}, {Oesch}, {Schouws}, {Stark}, {Stefanon}, {Aravena}, {da Cunha}, {De Looze}, {Fudamoto}, {Graziani}, {Hodge}, {Riechers}, {Schneider}, {Algera}, {Barrufet}, {Hygate}, {Labb{\'e}}, {Li}, {Nanayakkara}, {Topping}, \& {van der Werf}}]{Ferrara22}
{Ferrara}, A., {Sommovigo}, L., {Dayal}, P., {et~al.} 2022, \mnras, 512, 58

\bibitem[{{Feruglio} {et~al.}(2023){Feruglio}, {Maio}, {Tripodi}, {Winters}, {Zappacosta}, {Bischetti}, {Civano}, {Carniani}, {D'Odorico}, {Fiore}, {Gallerani}, {Ginolfi}, {Maiolino}, {Piconcelli}, {Valiante}, \& {Zanchettin}}]{Feruglio23}
{Feruglio}, C., {Maio}, U., {Tripodi}, R., {et~al.} 2023, \apjl, 954, L10

\bibitem[{{Fixsen} {et~al.}(1998){Fixsen}, {Dwek}, {Mather}, {Bennett}, \& {Shafer}}]{Fixsen98}
{Fixsen}, D.~J., {Dwek}, E., {Mather}, J.~C., {Bennett}, C.~L., \& {Shafer}, R.~A. 1998, \apj, 508, 123

\bibitem[{{Fujimoto} {et~al.}(2021){Fujimoto}, {Oguri}, {Brammer}, {Yoshimura}, {Laporte}, {Gonz{\'a}lez-L{\'o}pez}, {Caminha}, {Kohno}, {Zitrin}, {Richard}, {Ouchi}, {Bauer}, {Smail}, {Hatsukade}, {Ono}, {Kokorev}, {Umehata}, {Schaerer}, {Knudsen}, {Sun}, {Magdis}, {Valentino}, {Ao}, {Toft}, {Dessauges-Zavadsky}, {Shimasaku}, {Caputi}, {Kusakabe}, {Morokuma-Matsui}, {Shotaro}, {Egami}, {Lee}, {Rawle}, \& {Espada}}]{Fujimoto21}
{Fujimoto}, S., {Oguri}, M., {Brammer}, G., {et~al.} 2021, \apj, 911, 99

\bibitem[{{Fujimoto} {et~al.}(2019){Fujimoto}, {Ouchi}, {Ferrara}, {Pallottini}, {Ivison}, {Behrens}, {Gallerani}, {Arata}, {Yajima}, \& {Nagamine}}]{Fujimoto19}
{Fujimoto}, S., {Ouchi}, M., {Ferrara}, A., {et~al.} 2019, \apj, 887, 107

\bibitem[{{Fujimoto} {et~al.}(2024){Fujimoto}, {Ouchi}, {Nakajima}, {Harikane}, {Isobe}, {Brammer}, {Oguri}, {Gim{\'e}nez-Arteaga}, {Heintz}, {Kokorev}, {Bauer}, {Ferrara}, {Kojima}, {Lagos}, {Laura}, {Schaerer}, {Shimasaku}, {Hatsukade}, {Kohno}, {Sun}, {Valentino}, {Watson}, {Fudamoto}, {Inoue}, {Gonz{\'a}lez-L{\'o}pez}, {Koekemoer}, {Knudsen}, {Lee}, {Magdis}, {Richard}, {Strait}, {Sugahara}, {Tamura}, {Toft}, {Umehata}, \& {Walth}}]{Fujimoto24}
{Fujimoto}, S., {Ouchi}, M., {Nakajima}, K., {et~al.} 2024, \apj, 964, 146

\bibitem[{{Garcia} {et~al.}(2023){Garcia}, {Narayanan}, {Popping}, {Anirudh}, {Sutherland}, \& {Kaasinen}}]{Garcia23}
{Garcia}, K., {Narayanan}, D., {Popping}, G., {et~al.} 2023, arXiv e-prints, arXiv:2311.01508

\bibitem[{{Ginolfi} {et~al.}(2020){Ginolfi}, {Jones}, {B{\'e}thermin}, {Faisst}, {Lemaux}, {Schaerer}, {Fudamoto}, {Oesch}, {Dessauges-Zavadsky}, {Fujimoto}, {Carniani}, {Le F{\`e}vre}, {Cassata}, {Silverman}, {Capak}, {Yan}, {Bardelli}, {Cucciati}, {Gal}, {Gruppioni}, {Hathi}, {Lubin}, {Maiolino}, {Morselli}, {Pelliccia}, {Talia}, {Vergani}, \& {Zamorani}}]{Ginolfi20}
{Ginolfi}, M., {Jones}, G.~C., {B{\'e}thermin}, M., {et~al.} 2020, \aap, 643, A7

\bibitem[{{Gkogkou} {et~al.}(2023){Gkogkou}, {B{\'e}thermin}, {Lagache}, {Van Cuyck}, {Jullo}, {Aravena}, {Beelen}, {Benoit}, {Bounmy}, {Calvo}, {Catalano}, {Cora}, {Croton}, {de la Torre}, {Fasano}, {Ferrara}, {Goupy}, {Hoarau}, {Hu}, {Ishiyama}, {Knudsen}, {Lambert}, {Mac{\'\i}as-P{\'e}rez}, {Marpaud}, {Mellema}, {Monfardini}, {Pallottini}, {Ponthieu}, {Prada}, {Roehlly}, {Vallini}, \& {Walter}}]{Gkogkou23}
{Gkogkou}, A., {B{\'e}thermin}, M., {Lagache}, G., {et~al.} 2023, \aap, 670, A16

\bibitem[{{Glover} \& {Clark}(2016)}]{Glover16}
{Glover}, S. C.~O. \& {Clark}, P.~C. 2016, \mnras, 456, 3596

\bibitem[{{Goldsmith} {et~al.}(2012){Goldsmith}, {Langer}, {Pineda}, \& {Velusamy}}]{Goldsmith12}
{Goldsmith}, P.~F., {Langer}, W.~D., {Pineda}, J.~L., \& {Velusamy}, T. 2012, \apjs, 203, 13

\bibitem[{{Haardt} \& {Madau}(1996)}]{HaardtMadau96}
{Haardt}, F. \& {Madau}, P. 1996, \apj, 461, 20

\bibitem[{{Habing}(1968)}]{Habing68}
{Habing}, H.~J. 1968, \bain, 19, 421

\bibitem[{{Hagedorn} {et~al.}(2024){Hagedorn}, {Cicone}, {Sarzi}, {Saintonge}, {Severgnini}, {Vignali}, {Shen}, {Rubinur}, {Schimek}, \& {Lasrado}}]{Hagedorn24}
{Hagedorn}, B., {Cicone}, C., {Sarzi}, M., {et~al.} 2024, arXiv e-prints, arXiv:2402.18291

\bibitem[{{Harikane} {et~al.}(2024){Harikane}, {Nakajima}, {Ouchi}, {Umeda}, {Isobe}, {Ono}, {Xu}, \& {Zhang}}]{Harikane24}
{Harikane}, Y., {Nakajima}, K., {Ouchi}, M., {et~al.} 2024, \apj, 960, 56

\bibitem[{{Harikane} {et~al.}(2022){Harikane}, {Ono}, {Ouchi}, {Liu}, {Sawicki}, {Shibuya}, {Behroozi}, {He}, {Shimasaku}, {Arnouts}, {Coupon}, {Fujimoto}, {Gwyn}, {Huang}, {Inoue}, {Kashikawa}, {Komiyama}, {Matsuoka}, \& {Willott}}]{Harikane22}
{Harikane}, Y., {Ono}, Y., {Ouchi}, M., {et~al.} 2022, \apjs, 259, 20

\bibitem[{{Harikane} {et~al.}(2020){Harikane}, {Ouchi}, {Inoue}, {Matsuoka}, {Tamura}, {Bakx}, {Fujimoto}, {Moriwaki}, {Ono}, {Nagao}, {Tadaki}, {Kojima}, {Shibuya}, {Egami}, {Ferrara}, {Gallerani}, {Hashimoto}, {Kohno}, {Matsuda}, {Matsuo}, {Pallottini}, {Sugahara}, \& {Vallini}}]{Harikane20}
{Harikane}, Y., {Ouchi}, M., {Inoue}, A.~K., {et~al.} 2020, \apj, 896, 93

\bibitem[{{Harikane} {et~al.}(2023){Harikane}, {Ouchi}, {Oguri}, {Ono}, {Nakajima}, {Isobe}, {Umeda}, {Mawatari}, \& {Zhang}}]{Harikane23}
{Harikane}, Y., {Ouchi}, M., {Oguri}, M., {et~al.} 2023, \apjs, 265, 5

\bibitem[{{Hernandez-Monteagudo} {et~al.}(2017){Hernandez-Monteagudo}, {Maio}, {Ciardi}, \& {Sunyaev}}]{HM17}
{Hernandez-Monteagudo}, C., {Maio}, U., {Ciardi}, B., \& {Sunyaev}, R.~A. 2017, arXiv e-prints, arXiv:1707.01910

\bibitem[{{Herrera-Camus} {et~al.}(2015){Herrera-Camus}, {Bolatto}, {Wolfire}, {Smith}, {Croxall}, {Kennicutt}, {Calzetti}, {Helou}, {Walter}, {Leroy}, {Draine}, {Brandl}, {Armus}, {Sandstrom}, {Dale}, {Aniano}, {Meidt}, {Boquien}, {Hunt}, {Galametz}, {Tabatabaei}, {Murphy}, {Appleton}, {Roussel}, {Engelbracht}, \& {Beirao}}]{Herrera-Camus15}
{Herrera-Camus}, R., {Bolatto}, A.~D., {Wolfire}, M.~G., {et~al.} 2015, \apj, 800, 1

\bibitem[{{Hollenbach} \& {McKee}(1989)}]{Hollenbach89}
{Hollenbach}, D. \& {McKee}, C.~F. 1989, \apj, 342, 306

\bibitem[{{Hunt} {et~al.}(2020){Hunt}, {Tortora}, {Ginolfi}, \& {Schneider}}]{Hunt20}
{Hunt}, L.~K., {Tortora}, C., {Ginolfi}, M., \& {Schneider}, R. 2020, \aap, 643, A180

\bibitem[{{Jones} {et~al.}(2024){Jones}, {Witstok}, {Concas}, \& {Laporte}}]{Jones24}
{Jones}, G.~C., {Witstok}, J., {Concas}, A., \& {Laporte}, N. 2024, \mnras, 529, L1

\bibitem[{{Kaasinen} {et~al.}(2024){Kaasinen}, {Venemans}, {Harrington}, {Boogaard}, {Meyer}, {Ba{\~n}ados}, {Decarli}, {Walter}, {Neeleman}, {Rivera}, \& {da Cunha}}]{Kaasinen24}
{Kaasinen}, M., {Venemans}, B., {Harrington}, K.~C., {et~al.} 2024, \aap, 684, A33

\bibitem[{{Kannan} {et~al.}(2022){Kannan}, {Smith}, {Garaldi}, {Shen}, {Vogelsberger}, {Pakmor}, {Springel}, \& {Hernquist}}]{Kannan22}
{Kannan}, R., {Smith}, A., {Garaldi}, E., {et~al.} 2022, \mnras, 514, 3857

\bibitem[{{Katz} {et~al.}(2019){Katz}, {Galligan}, {Kimm}, {Rosdahl}, {Haehnelt}, {Blaizot}, {Devriendt}, {Slyz}, {Laporte}, \& {Ellis}}]{Katz19}
{Katz}, H., {Galligan}, T.~P., {Kimm}, T., {et~al.} 2019, \mnras, 487, 5902

\bibitem[{{Katz} {et~al.}(2023){Katz}, {Saxena}, {Rosdahl}, {Kimm}, {Blaizot}, {Garel}, {Michel-Dansac}, {Haehnelt}, {Ellis}, {Pentericci}, {Devriendt}, \& {Slyz}}]{Katz23}
{Katz}, H., {Saxena}, A., {Rosdahl}, J., {et~al.} 2023, \mnras, 518, 270

\bibitem[{{Khusanova} {et~al.}(2021){Khusanova}, {Bethermin}, {Le F{\`e}vre}, {Capak}, {Faisst}, {Schaerer}, {Silverman}, {Cassata}, {Yan}, {Ginolfi}, {Fudamoto}, {Loiacono}, {Amorin}, {Bardelli}, {Boquien}, {Cimatti}, {Dessauges-Zavadsky}, {Gruppioni}, {Hathi}, {Jones}, {Koekemoer}, {Lagache}, {Maiolino}, {Lemaux}, {Oesch}, {Pozzi}, {Riechers}, {Romano}, {Talia}, {Toft}, {Vergani}, {Zamorani}, \& {Zucca}}]{Khusanova21}
{Khusanova}, Y., {Bethermin}, M., {Le F{\`e}vre}, O., {et~al.} 2021, \aap, 649, A152

\bibitem[{{Klaassen} {et~al.}(2020){Klaassen}, {Mroczkowski}, {Cicone}, {Hatziminaoglou}, {Sartori}, {De Breuck}, {Bryan}, {Dicker}, {Duran}, {Groppi}, {Kaercher}, {Kawabe}, {Kohno}, \& {Geach}}]{Klaassen20}
{Klaassen}, P.~D., {Mroczkowski}, T.~K., {Cicone}, C., {et~al.} 2020, in Society of Photo-Optical Instrumentation Engineers (SPIE) Conference Series, Vol. 11445, Ground-based and Airborne Telescopes VIII, ed. H.~K. {Marshall}, J.~{Spyromilio}, \& T.~{Usuda}, 114452F

\bibitem[{{Kobayashi} {et~al.}(2006){Kobayashi}, {Umeda}, {Nomoto}, {Tominaga}, \& {Ohkubo}}]{Kobayashi06}
{Kobayashi}, C., {Umeda}, H., {Nomoto}, K., {Tominaga}, N., \& {Ohkubo}, T. 2006, \apj, 653, 1145

\bibitem[{{Krumholz}(2013)}]{Krumholz13}
{Krumholz}, M.~R. 2013, \mnras, 436, 2747

\bibitem[{{Lagache} {et~al.}(2018){Lagache}, {Cousin}, \& {Chatzikos}}]{Lagache18}
{Lagache}, G., {Cousin}, M., \& {Chatzikos}, M. 2018, \aap, 609, A130

\bibitem[{{Lagos} {et~al.}(2015){Lagos}, {Crain}, {Schaye}, {Furlong}, {Frenk}, {Bower}, {Schaller}, {Theuns}, {Trayford}, {Bah{\'e}}, \& {Dalla Vecchia}}]{Lagos15}
{Lagos}, C. d.~P., {Crain}, R.~A., {Schaye}, J., {et~al.} 2015, \mnras, 452, 3815

\bibitem[{{Lah{\'e}n} {et~al.}(2024){Lah{\'e}n}, {Naab}, \& {Sz{\'e}csi}}]{Lahen24}
{Lah{\'e}n}, N., {Naab}, T., \& {Sz{\'e}csi}, D. 2024, \mnras, 530, 645

\bibitem[{{Laporte} {et~al.}(2019){Laporte}, {Katz}, {Ellis}, {Lagache}, {Bauer}, {Boone}, {Inoue}, {Hashimoto}, {Matsuo}, {Mawatari}, \& {Tamura}}]{Laporte19}
{Laporte}, N., {Katz}, H., {Ellis}, R.~S., {et~al.} 2019, \mnras, 487, L81

\bibitem[{{Le F{\`e}vre} {et~al.}(2020){Le F{\`e}vre}, {B{\'e}thermin}, {Faisst}, {Jones}, {Capak}, {Cassata}, {Silverman}, {Schaerer}, {Yan}, {Amorin}, {Bardelli}, {Boquien}, {Cimatti}, {Dessauges-Zavadsky}, {Giavalisco}, {Hathi}, {Fudamoto}, {Fujimoto}, {Ginolfi}, {Gruppioni}, {Hemmati}, {Ibar}, {Koekemoer}, {Khusanova}, {Lagache}, {Lemaux}, {Loiacono}, {Maiolino}, {Mancini}, {Narayanan}, {Morselli}, {M{\'e}ndez-Hern{\`a}ndez}, {Oesch}, {Pozzi}, {Romano}, {Riechers}, {Scoville}, {Talia}, {Tasca}, {Thomas}, {Toft}, {Vallini}, {Vergani}, {Walter}, {Zamorani}, \& {Zucca}}]{LaFevre20}
{Le F{\`e}vre}, O., {B{\'e}thermin}, M., {Faisst}, A., {et~al.} 2020, \aap, 643, A1

\bibitem[{{Leech} {et~al.}(1999){Leech}, {V{\"o}lk}, {Heinrichsen}, {Hippelein}, {Metcalfe}, {Pierini}, {Popescu}, {Tuffs}, \& {Xu}}]{Leech99}
{Leech}, K.~J., {V{\"o}lk}, H.~J., {Heinrichsen}, I., {et~al.} 1999, \mnras, 310, 317

\bibitem[{{Leethochawalit} {et~al.}(2023{\natexlab{a}}){Leethochawalit}, {Roberts-Borsani}, {Morishita}, {Trenti}, \& {Treu}}]{Leethochawalit22b}
{Leethochawalit}, N., {Roberts-Borsani}, G., {Morishita}, T., {Trenti}, M., \& {Treu}, T. 2023{\natexlab{a}}, \mnras, 524, 5454

\bibitem[{{Leethochawalit} {et~al.}(2023{\natexlab{b}}){Leethochawalit}, {Trenti}, {Santini}, {Yang}, {Merlin}, {Castellano}, {Fontana}, {Treu}, {Mason}, {Glazebrook}, {Jones}, {Vulcani}, {Nanayakkara}, {Marchesini}, {Mascia}, {Morishita}, {Roberts-Borsani}, {Bonchi}, {Paris}, {Boyett}, {Strait}, {Calabr{\`o}}, {Pentericci}, {Bradac}, {Wang}, \& {Scarlata}}]{Leethochawalit22a}
{Leethochawalit}, N., {Trenti}, M., {Santini}, P., {et~al.} 2023{\natexlab{b}}, \apjl, 942, L26

\bibitem[{{Leung} {et~al.}(2020){Leung}, {Olsen}, {Somerville}, {Dav{\'e}}, {Greve}, {Hayward}, {Narayanan}, \& {Popping}}]{Leung20}
{Leung}, T.~K.~D., {Olsen}, K.~P., {Somerville}, R.~S., {et~al.} 2020, \apj, 905, 102

\bibitem[{{Liang} {et~al.}(2024){Liang}, {Feldmann}, {Murray}, {Narayanan}, {Hayward}, {Angl{\'e}s-Alc{\'a}zar}, {Bassini}, {Richings}, {Faucher-Gigu{\`e}re}, {Chung}, {Chan}, {Tolgay}, {{\c{C}}atmabacak}, {Kere{\v{s}}}, \& {Hopkins}}]{Liang24}
{Liang}, L., {Feldmann}, R., {Murray}, N., {et~al.} 2024, \mnras, 528, 499

\bibitem[{{Limongi} \& {Chieffi}(2018)}]{Limongi18}
{Limongi}, M. \& {Chieffi}, A. 2018, \apjs, 237, 13

\bibitem[{{Ma} {et~al.}(2017{\natexlab{a}}){Ma}, {Maio}, {Ciardi}, \& {Salvaterra}}]{Ma17a}
{Ma}, Q., {Maio}, U., {Ciardi}, B., \& {Salvaterra}, R. 2017{\natexlab{a}}, \mnras, 466, 1140

\bibitem[{{Ma} {et~al.}(2017{\natexlab{b}}){Ma}, {Maio}, {Ciardi}, \& {Salvaterra}}]{Ma17b}
{Ma}, Q., {Maio}, U., {Ciardi}, B., \& {Salvaterra}, R. 2017{\natexlab{b}}, \mnras, 472, 3532

\bibitem[{{Madden} {et~al.}(2020){Madden}, {Cormier}, {Hony}, {Lebouteiller}, {Abel}, {Galametz}, {De Looze}, {Chevance}, {Polles}, {Lee}, {Galliano}, {Lambert-Huyghe}, {Hu}, \& {Ramambason}}]{Madden20}
{Madden}, S.~C., {Cormier}, D., {Hony}, S., {et~al.} 2020, \aap, 643, A141

\bibitem[{{Maio} {et~al.}(2010){Maio}, {Ciardi}, {Dolag}, {Tornatore}, \& {Khochfar}}]{Maio10}
{Maio}, U., {Ciardi}, B., {Dolag}, K., {Tornatore}, L., \& {Khochfar}, S. 2010, \mnras, 407, 1003

\bibitem[{{Maio} {et~al.}(2007){Maio}, {Dolag}, {Ciardi}, \& {Tornatore}}]{Maio07}
{Maio}, U., {Dolag}, K., {Ciardi}, B., \& {Tornatore}, L. 2007, \mnras, 379, 963

\bibitem[{{Maio} {et~al.}(2011){Maio}, {Khochfar}, {Johnson}, \& {Ciardi}}]{Maio11}
{Maio}, U., {Khochfar}, S., {Johnson}, J.~L., \& {Ciardi}, B. 2011, \mnras, 414, 1145

\bibitem[{{Maio} {et~al.}(2022){Maio}, {P{\'e}roux}, \& {Ciardi}}]{Maio22}
{Maio}, U., {P{\'e}roux}, C., \& {Ciardi}, B. 2022, \aap, 657, A47

\bibitem[{{Maio} {et~al.}(2016){Maio}, {Petkova}, {De Lucia}, \& {Borgani}}]{Maio16}
{Maio}, U., {Petkova}, M., {De Lucia}, G., \& {Borgani}, S. 2016, \mnras, 460, 3733

\bibitem[{{Maio} \& {Tescari}(2015)}]{Maio15}
{Maio}, U. \& {Tescari}, E. 2015, \mnras, 453, 3798

\bibitem[{{Maio} \& {Viel}(2023)}]{Maio23}
{Maio}, U. \& {Viel}, M. 2023, \aap, 672, A71

\bibitem[{{Maiolino} {et~al.}(2015){Maiolino}, {Carniani}, {Fontana}, {Vallini}, {Pentericci}, {Ferrara}, {Vanzella}, {Grazian}, {Gallerani}, {Castellano}, {Cristiani}, {Brammer}, {Santini}, {Wagg}, \& {Williams}}]{Maiolino15}
{Maiolino}, R., {Carniani}, S., {Fontana}, A., {et~al.} 2015, \mnras, 452, 54

\bibitem[{{McElroy} {et~al.}(2013){McElroy}, {Walsh}, {Markwick}, {Cordiner}, {Smith}, \& {Millar}}]{McElroy13}
{McElroy}, D., {Walsh}, C., {Markwick}, A.~J., {et~al.} 2013, \aap, 550, A36

\bibitem[{{Merlin} {et~al.}(2019){Merlin}, {Fortuni}, {Torelli}, {Santini}, {Castellano}, {Fontana}, {Grazian}, {Pentericci}, {Pilo}, \& {Schmidt}}]{Merlin19}
{Merlin}, E., {Fortuni}, F., {Torelli}, M., {et~al.} 2019, \mnras, 490, 3309

\bibitem[{{Nakajima} {et~al.}(2023){Nakajima}, {Ouchi}, {Isobe}, {Harikane}, {Zhang}, {Ono}, {Umeda}, \& {Oguri}}]{Nakajima23}
{Nakajima}, K., {Ouchi}, M., {Isobe}, Y., {et~al.} 2023, \apjs, 269, 33

\bibitem[{{Narayanan} \& {Krumholz}(2017)}]{Narayanan17}
{Narayanan}, D. \& {Krumholz}, M.~R. 2017, \mnras, 467, 50

\bibitem[{{Nordon} \& {Sternberg}(2016)}]{Nordon16}
{Nordon}, R. \& {Sternberg}, A. 2016, \mnras, 462, 2804

\bibitem[{{Olsen} {et~al.}(2017){Olsen}, {Greve}, {Narayanan}, {Thompson}, {Dav{\'e}}, {Niebla Rios}, \& {Stawinski}}]{Olsen17}
{Olsen}, K., {Greve}, T.~R., {Narayanan}, D., {et~al.} 2017, \apj, 846, 105

\bibitem[{{Padmanabhan}(2019)}]{Padma19}
{Padmanabhan}, H. 2019, \mnras, 488, 3014

\bibitem[{{Pallottini} {et~al.}(2019){Pallottini}, {Ferrara}, {Decataldo}, {Gallerani}, {Vallini}, {Carniani}, {Behrens}, {Kohandel}, \& {Salvadori}}]{Pallottini19}
{Pallottini}, A., {Ferrara}, A., {Decataldo}, D., {et~al.} 2019, \mnras, 487, 1689

\bibitem[{{Pentericci} {et~al.}(2016){Pentericci}, {Carniani}, {Castellano}, {Fontana}, {Maiolino}, {Guaita}, {Vanzella}, {Grazian}, {Santini}, {Yan}, {Cristiani}, {Conselice}, {Giavalisco}, {Hathi}, \& {Koekemoer}}]{Pentericci16}
{Pentericci}, L., {Carniani}, S., {Castellano}, M., {et~al.} 2016, \apjl, 829, L11

\bibitem[{{P{\'e}roux} \& {Howk}(2020)}]{Peroux20}
{P{\'e}roux}, C. \& {Howk}, J.~C. 2020, \araa, 58, 363

\bibitem[{{Pettini} {et~al.}(2003){Pettini}, {Madau}, {Bolte}, {Prochaska}, {Ellison}, \& {Fan}}]{Pettini03}
{Pettini}, M., {Madau}, P., {Bolte}, M., {et~al.} 2003, \apj, 594, 695

\bibitem[{{Pineda} {et~al.}(2013){Pineda}, {Langer}, {Velusamy}, \& {Goldsmith}}]{Pineda13}
{Pineda}, J.~L., {Langer}, W.~D., {Velusamy}, T., \& {Goldsmith}, P.~F. 2013, \aap, 554, A103

\bibitem[{{Pizzati} {et~al.}(2020){Pizzati}, {Ferrara}, {Pallottini}, {Gallerani}, {Vallini}, {Decataldo}, \& {Fujimoto}}]{Pizzati20}
{Pizzati}, E., {Ferrara}, A., {Pallottini}, A., {et~al.} 2020, \mnras, 495, 160

\bibitem[{{Pizzati} {et~al.}(2023){Pizzati}, {Ferrara}, {Pallottini}, {Sommovigo}, {Kohandel}, \& {Carniani}}]{Pizzati23}
{Pizzati}, E., {Ferrara}, A., {Pallottini}, A., {et~al.} 2023, \mnras, 519, 4608

\bibitem[{{Popping} {et~al.}(2019){Popping}, {Narayanan}, {Somerville}, {Faisst}, \& {Krumholz}}]{Popping19}
{Popping}, G., {Narayanan}, D., {Somerville}, R.~S., {Faisst}, A.~L., \& {Krumholz}, M.~R. 2019, \mnras, 482, 4906

\bibitem[{{Popping} \& {P{\'e}roux}(2022)}]{Popping22}
{Popping}, G. \& {P{\'e}roux}, C. 2022, \mnras, 513, 1531

\bibitem[{{Posses} {et~al.}(2024){Posses}, {Aravena}, {Gonz{\'a}lez-L{\'o}pez}, {F{\"o}rster Schreiber}, {Liu}, {Lee}, {Solimano}, {D{\'\i}az-Santos}, {Assef}, {Barcos-Mu{\~n}oz}, {Bovino}, {Bowler}, {Calistro Rivera}, {da Cunha}, {Davies}, {Killi}, {De Looze}, {Ferrara}, {Fisher}, {Herrera-Camus}, {Ikeda}, {Lambert}, {Li}, {Lutz}, {Mitsuhashi}, {Palla}, {Rela{\~n}o}, {Spilker}, {Naab}, {Tadaki}, {Telikova}, {{\"U}bler}, {van der Giessen}, \& {Villanueva}}]{Posses24}
{Posses}, A., {Aravena}, M., {Gonz{\'a}lez-L{\'o}pez}, J., {et~al.} 2024, arXiv e-prints, arXiv:2403.03379

\bibitem[{{Pozzi} {et~al.}(2024){Pozzi}, {Calura}, {D'Amato}, {Gavarente}, {Bethermin}, {Boquien}, {Casasola}, {Cimatti}, {Cochrane}, {Dessauges-Zavadsky}, {Enia}, {Esposito}, {Faisst}, {Gilli}, {Ginolfi}, {Gobat}, {Gruppioni}, {Hayward}, {Ibar}, {Koekemoer}, {Lemaux}, {Magdis}, {Molina}, {Romano}, {Talia}, {Vallini}, {Vergani}, \& {Zamorani}}]{Pozzi24}
{Pozzi}, F., {Calura}, F., {D'Amato}, Q., {et~al.} 2024, \aap, 686, A187

\bibitem[{{Rigopoulou} {et~al.}(2014){Rigopoulou}, {Hopwood}, {Magdis}, {Thatte}, {Swinyard}, {Farrah}, {Huang}, {Alonso-Herrero}, {Bock}, {Clements}, {Cooray}, {Griffin}, {Oliver}, {Pearson}, {Riechers}, {Scott}, {Smith}, {Vaccari}, {Valtchanov}, \& {Wang}}]{Rigopoulou14}
{Rigopoulou}, D., {Hopwood}, R., {Magdis}, G.~E., {et~al.} 2014, \apjl, 781, L15

\bibitem[{{Romano} {et~al.}(2024){Romano}, {Donevski}, {Junais}, {Nanni}, {Ginolfi}, {Jones}, {Shivaei}, {Lorenzon}, {Hamed}, {Salak}, \& {Sawant}}]{Romano24}
{Romano}, M., {Donevski}, D., {Junais}, {et~al.} 2024, \aap, 683, L9

\bibitem[{{Romano} {et~al.}(2022){Romano}, {Morselli}, {Cassata}, {Ginolfi}, {Schaerer}, {B{\'e}thermin}, {Capak}, {Faisst}, {Le F{\`e}vre}, {Silverman}, {Yan}, {Bardelli}, {Boquien}, {Dessauges-Zavadsky}, {Fujimoto}, {Hathi}, {Jones}, {Koekemoer}, {Lemaux}, {M{\'e}ndez-Hern{\'a}ndez}, {Narayanan}, {Talia}, {Vergani}, {Zamorani}, \& {Zucca}}]{Romano22}
{Romano}, M., {Morselli}, L., {Cassata}, P., {et~al.} 2022, \aap, 660, A14

\bibitem[{{Rowland} {et~al.}(2024){Rowland}, {Hodge}, {Bouwens}, {Mancera Pi{\~n}a}, {Hygate}, {Algera}, {Aravena}, {Bowler}, {da Cunha}, {Dayal}, {Ferrara}, {Herard-Demanche}, {Inami}, {van Leeuwen}, {de Looze}, {Oesch}, {Pallottini}, {Phillips}, {Rybak}, {Schouws}, {Smit}, {Sommovigo}, {Stefanon}, \& {van der Werf}}]{Rowland24}
{Rowland}, L.~E., {Hodge}, J., {Bouwens}, R., {et~al.} 2024, arXiv e-prints, arXiv:2405.06025

\bibitem[{{Ryan-Weber} {et~al.}(2009){Ryan-Weber}, {Pettini}, {Madau}, \& {Zych}}]{Ryan-Weber09}
{Ryan-Weber}, E.~V., {Pettini}, M., {Madau}, P., \& {Zych}, B.~J. 2009, \mnras, 395, 1476

\bibitem[{{Saintonge} {et~al.}(2017){Saintonge}, {Catinella}, {Tacconi}, {Kauffmann}, {Genzel}, {Cortese}, {Dav{\'e}}, {Fletcher}, {Graci{\'a}-Carpio}, {Kramer}, {Heckman}, {Janowiecki}, {Lutz}, {Rosario}, {Schiminovich}, {Schuster}, {Wang}, {Wuyts}, {Borthakur}, {Lamperti}, \& {Roberts-Borsani}}]{Saintonge17}
{Saintonge}, A., {Catinella}, B., {Tacconi}, L.~J., {et~al.} 2017, \apjs, 233, 22

\bibitem[{{Salim} \& {Narayanan}(2020)}]{Salim20}
{Salim}, S. \& {Narayanan}, D. 2020, \araa, 58, 529

\bibitem[{{Santoro} \& {Shull}(2006)}]{Santoro06}
{Santoro}, F. \& {Shull}, J.~M. 2006, \apj, 643, 26

\bibitem[{{Sargent} {et~al.}(1988){Sargent}, {Boksenberg}, \& {Steidel}}]{Sargent88}
{Sargent}, W. L.~W., {Boksenberg}, A., \& {Steidel}, C.~C. 1988, \apjs, 68, 539

\bibitem[{{Sargsyan} {et~al.}(2012){Sargsyan}, {Lebouteiller}, {Weedman}, {Spoon}, {Bernard-Salas}, {Engels}, {Stacey}, {Houck}, {Barry}, {Miles}, \& {Samsonyan}}]{Sargsyan12}
{Sargsyan}, L., {Lebouteiller}, V., {Weedman}, D., {et~al.} 2012, \apj, 755, 171

\bibitem[{{Schaerer} {et~al.}(2020){Schaerer}, {Ginolfi}, {B{\'e}thermin}, {Fudamoto}, {Oesch}, {Le F{\`e}vre}, {Faisst}, {Capak}, {Cassata}, {Silverman}, {Yan}, {Jones}, {Amorin}, {Bardelli}, {Boquien}, {Cimatti}, {Dessauges-Zavadsky}, {Giavalisco}, {Hathi}, {Fujimoto}, {Ibar}, {Koekemoer}, {Lagache}, {Lemaux}, {Loiacono}, {Maiolino}, {Narayanan}, {Morselli}, {M{\'e}ndez-Hern{\`a}ndez}, {Pozzi}, {Riechers}, {Talia}, {Toft}, {Vallini}, {Vergani}, {Zamorani}, \& {Zucca}}]{Schaerer20}
{Schaerer}, D., {Ginolfi}, M., {B{\'e}thermin}, M., {et~al.} 2020, \aap, 643, A3

\bibitem[{{Schimek} {et~al.}(2024){Schimek}, {Decataldo}, {Shen}, {Cicone}, {Baumschlager}, {van Kampen}, {Klaassen}, {Madau}, {Di Mascolo}, {Mayer}, {Montoya Arroyave}, {Mroczkowski}, \& {Warraich}}]{Schimek24}
{Schimek}, A., {Decataldo}, D., {Shen}, S., {et~al.} 2024, \aap, 682, A98

\bibitem[{{Schouws} {et~al.}(2022){Schouws}, {Stefanon}, {Bouwens}, {Smit}, {Hodge}, {Labb{\'e}}, {Algera}, {Boogaard}, {Carniani}, {Fudamoto}, {Holwerda}, {Illingworth}, {Maiolino}, {Maseda}, {Oesch}, \& {van der Werf}}]{Schouws22}
{Schouws}, S., {Stefanon}, M., {Bouwens}, R., {et~al.} 2022, \apj, 928, 31

\bibitem[{{Sebastian} {et~al.}(2024){Sebastian}, {Ryan-Weber}, {Davies}, {Becker}, {Keating}, {D'Odorico}, {Meyer}, {Bosman}, {Cupani}, {Kulkarni}, {Haehnelt}, {Lai}, {Eilers}, {Bischetti}, \& {Gallerani}}]{Sebastian24}
{Sebastian}, A.~M., {Ryan-Weber}, E., {Davies}, R.~L., {et~al.} 2024, \mnras, 530, 1829

\bibitem[{{Simcoe} {et~al.}(2011){Simcoe}, {Cooksey}, {Matejek}, {Burgasser}, {Bochanski}, {Lovegrove}, {Bernstein}, {Pipher}, {Forrest}, {McMurtry}, {Fan}, \& {O'Meara}}]{Simcoe11}
{Simcoe}, R.~A., {Cooksey}, K.~L., {Matejek}, M., {et~al.} 2011, \apj, 743, 21

\bibitem[{{Simcoe} {et~al.}(2020){Simcoe}, {Onoue}, {Eilers}, {Banados}, {Cooper}, {Furesz}, {Hennawi}, \& {Venemans}}]{Simcoe20}
{Simcoe}, R.~A., {Onoue}, M., {Eilers}, A.-C., {et~al.} 2020, arXiv e-prints, arXiv:2011.10582

\bibitem[{{Simcoe} {et~al.}(2006){Simcoe}, {Sargent}, {Rauch}, \& {Becker}}]{Simcoe06}
{Simcoe}, R.~A., {Sargent}, W. L.~W., {Rauch}, M., \& {Becker}, G. 2006, \apj, 637, 648

\bibitem[{{Sommovigo} {et~al.}(2021){Sommovigo}, {Ferrara}, {Carniani}, {Zanella}, {Pallottini}, {Gallerani}, \& {Vallini}}]{Sommovigo21}
{Sommovigo}, L., {Ferrara}, A., {Carniani}, S., {et~al.} 2021, \mnras, 503, 4878

\bibitem[{{Sommovigo} {et~al.}(2022){Sommovigo}, {Ferrara}, {Pallottini}, {Dayal}, {Bouwens}, {Smit}, {da Cunha}, {De Looze}, {Bowler}, {Hodge}, {Inami}, {Oesch}, {Endsley}, {Gonzalez}, {Schouws}, {Stark}, {Stefanon}, {Aravena}, {Graziani}, {Riechers}, {Schneider}, {van der Werf}, {Algera}, {Barrufet}, {Fudamoto}, {Hygate}, {Labb{\'e}}, {Li}, {Nanayakkara}, \& {Topping}}]{Sommovigo22}
{Sommovigo}, L., {Ferrara}, A., {Pallottini}, A., {et~al.} 2022, \mnras, 513, 3122

\bibitem[{{Songaila}(2001)}]{Songaila01}
{Songaila}, A. 2001, \apjl, 561, L153

\bibitem[{{Springel}(2005)}]{Springel05}
{Springel}, V. 2005, \mnras, 364, 1105

\bibitem[{{Springel} \& {Hernquist}(2003)}]{Springel03}
{Springel}, V. \& {Hernquist}, L. 2003, \mnras, 339, 289

\bibitem[{{Stacey} {et~al.}(1991){Stacey}, {Geis}, {Genzel}, {Lugten}, {Poglitsch}, {Sternberg}, \& {Townes}}]{Stacey91}
{Stacey}, G.~J., {Geis}, N., {Genzel}, R., {et~al.} 1991, \apj, 373, 423

\bibitem[{{Szakacs} {et~al.}(2022){Szakacs}, {P{\'e}roux}, {Zwaan}, {Nelson}, {Schinnerer}, {Lah{\'e}n}, {Weng}, \& {Fresco}}]{Szakacs22}
{Szakacs}, R., {P{\'e}roux}, C., {Zwaan}, M.~A., {et~al.} 2022, \mnras, 512, 4736

\bibitem[{{Tacchella} {et~al.}(2022){Tacchella}, {Finkelstein}, {Bagley}, {Dickinson}, {Ferguson}, {Giavalisco}, {Graziani}, {Grogin}, {Hathi}, {Hutchison}, {Jung}, {Koekemoer}, {Larson}, {Papovich}, {Pirzkal}, {Rojas-Ruiz}, {Song}, {Schneider}, {Somerville}, {Wilkins}, \& {Yung}}]{Tacchella22}
{Tacchella}, S., {Finkelstein}, S.~L., {Bagley}, M., {et~al.} 2022, \apj, 927, 170

\bibitem[{{Tescari} {et~al.}(2011){Tescari}, {Viel}, {D'Odorico}, {Cristiani}, {Calura}, {Borgani}, \& {Tornatore}}]{Tescari2011}
{Tescari}, E., {Viel}, M., {D'Odorico}, V., {et~al.} 2011, \mnras, 411, 826

\bibitem[{{Tumlinson} {et~al.}(2017){Tumlinson}, {Peeples}, \& {Werk}}]{Tumlinson17}
{Tumlinson}, J., {Peeples}, M.~S., \& {Werk}, J.~K. 2017, \araa, 55, 389

\bibitem[{{Velusamy} \& {Langer}(2014)}]{Velusamy14}
{Velusamy}, T. \& {Langer}, W.~D. 2014, \aap, 572, A45

\bibitem[{{Vizgan} {et~al.}(2022{\natexlab{a}}){Vizgan}, {Greve}, {Olsen}, {Zanella}, {Narayanan}, {Dav{\`e}}, {Magdis}, {Popping}, {Valentino}, \& {Heintz}}]{Vizgan22a}
{Vizgan}, D., {Greve}, T.~R., {Olsen}, K.~P., {et~al.} 2022{\natexlab{a}}, \apj, 929, 92

\bibitem[{{Vizgan} {et~al.}(2022{\natexlab{b}}){Vizgan}, {Heintz}, {Greve}, {Narayanan}, {Dav{\'e}}, {Olsen}, {Popping}, \& {Watson}}]{Vizgan22b}
{Vizgan}, D., {Heintz}, K.~E., {Greve}, T.~R., {et~al.} 2022{\natexlab{b}}, \apjl, 939, L1

\bibitem[{{Wheeler} {et~al.}(1995){Wheeler}, {Harkness}, {Khokhlov}, \& {Hoeflich}}]{Wheeler95}
{Wheeler}, J.~C., {Harkness}, R.~P., {Khokhlov}, A.~M., \& {Hoeflich}, P. 1995, \physrep, 256, 211

\bibitem[{{Whitaker} {et~al.}(2017){Whitaker}, {Pope}, {Cybulski}, {Casey}, {Popping}, \& {Yun}}]{Whitaker17}
{Whitaker}, K.~E., {Pope}, A., {Cybulski}, R., {et~al.} 2017, \apj, 850, 208

\bibitem[{{Wilson} \& {Bell}(2002)}]{Wilson02}
{Wilson}, N.~J. \& {Bell}, K.~L. 2002, \mnras, 337, 1027

\bibitem[{{Yoshida} {et~al.}(2003){Yoshida}, {Abel}, {Hernquist}, \& {Sugiyama}}]{Yoshida03}
{Yoshida}, N., {Abel}, T., {Hernquist}, L., \& {Sugiyama}, N. 2003, \apj, 592, 645

\bibitem[{{Zanella} {et~al.}(2018){Zanella}, {Daddi}, {Magdis}, {Diaz Santos}, {Cormier}, {Liu}, {Cibinel}, {Gobat}, {Dickinson}, {Sargent}, {Popping}, {Madden}, {Bethermin}, {Hughes}, {Valentino}, {Rujopakarn}, {Pannella}, {Bournaud}, {Walter}, {Wang}, {Elbaz}, \& {Coogan}}]{Zanella18}
{Zanella}, A., {Daddi}, E., {Magdis}, G., {et~al.} 2018, \mnras, 481, 1976

\bibitem[{{Zanella} {et~al.}(2023){Zanella}, {Valentino}, {Gallazzi}, {Belli}, {Magdis}, \& {Bolamperti}}]{Zanella23}
{Zanella}, A., {Valentino}, F., {Gallazzi}, A., {et~al.} 2023, \mnras, 524, 923

\end{thebibliography}

\section*{Appendix A: \CII Cooling}
\label{sect: appendix cii cooling}

In this Appendix, we provide the atomic data adopted in Section \ref{sect:LCII} to compute $\Lambda_{\textup{\CII}}$. We model \CII as a two-level system considering the fine structure transition 
$(2p)[\rm {}^2 P_{3/2} - {}^2 P_{1/2}]$ 
between the quantum number $J = 3/2$ and $J = 1/2$ states. The following data are taken from \cite{Hollenbach89} and \cite{Goldsmith12} and refer to 
$e^-$-impact, H-impact and H$_2$-impact collisional rates, spontaneous transition rate, and energy level separation:
\begin{equation*}
    \gamma ^{ { \rm e}}_{21} = 2.8\cdot 10^{-7} \, { T_{100}^{-0.5}}\; \; \; { \rm cm^3 \: s^{-1}},
\end{equation*}
\begin{equation*}
    \gamma ^{ { \rm H}}_{21} = 8\cdot 10^{-10} \, { T_{100}^{0.07}}\; \; \; {\rm cm^3 \: s^{-1}},
\end{equation*}
\begin{equation*}
    \gamma ^{ { \rm H_2}}_{21} = 3.8\cdot 10^{-10} \, { T_{100}^{0.14}}\; \; \; {\rm cm^3 \: s^{-1}},
\end{equation*}
\begin{equation*}
    A_{21} = 2.1 \cdot 10^{-4 } \; { \rm s^{-1}},
\end{equation*}
\begin{equation*}
    \Delta E_{21} = 5.71 \cdot 10^{-14 } \; {\rm  erg}.
\end{equation*}

\noindent 
where ${ T_{100}} =  T/(100$ K) and the validity regime is for gas with temperature $T < 10^4$ K.

\end{document}